\documentstyle[12pt]{article}

\global\arraycolsep=1pt
\newcommand{\CR}{\nonumber \\}

\def\be{\begin{equation}}
\def\ee{\end{equation}}
\def\bea{\begin{eqnarray}}
\def\eea{\end{eqnarray}}

\def\r{{\rm r}}
\def\rr{{\rm r}}

\def\grad{\vec{\nabla}}

\begin{document}

\renewcommand{\thefootnote}{\fnsymbol{footnote}}
\begin{titlepage}
\begin{flushright}
         CERN-9894, LPTHE-9818, NYU-9809\\
   hep-th/9807024\\
June 16, 1998
\end{flushright}
\begin{center}
{\Large \bf Renormalizable Non-Covariant Gauges\\and
Coulomb Gauge Limit }
\lineskip .75em
\vskip 2em
\normalsize
{\large  Laurent Baulieu}\footnote{email address:
baulieu@lpthe.jussieu.fr}   \\
{\it CERN, CH-1211 Gen\`{e}ve 23, Switzerland} \\
{\it LPTHE, Universit\'es Paris VI - Paris VII, Paris,
France}\footnote{
URA 280 CNRS,
4 place Jussieu, F-75252 Paris Cedex 05, France.} \\
{\it SISSA, Via Beirut.2-4, 34013 Trieste, Italia}

\vskip 1 em
{\large Daniel Zwanziger}\footnote{email address:
daniel.zwanziger@nyu.edu} 
\footnote{Work supported in part by National Science Foundation Grant no.
PHY9520978} \\ 
{\it  Department of Physics, New York University, New York 10003, USA}
\end{center}
\vskip 1 em
\begin{abstract} 
To study ``physical'' gauges such as the Coulomb, light-cone, axial or
temporal gauge, we consider ``interpolating'' gauges which interpolate
linearly between a covariant gauge, such as the Feynman or Landau gauge, and a
physical gauge.  Lorentz breaking by the gauge-fixing term of interpolating
gauges is controlled by extending the BRST method to include not only the
local gauge group, but also the global Lorentz group.  We enumerate the
possible divergences of interpolating gauges, and show that they are
renormalizable, and we show that the expectation value of physical
observables is the same as in a covariant gauge.  In the second part of the
article we study the Coulomb-gauge as the singular limit of the Landau-Coulomb
interpolating gauge.  We find that unrenormalized and renormalized
correlation functions are finite in this limit.  We also find that there are
finite two-loop diagrams of ``unphysical'' particles that are not present in
formal canonical quantization in the Coulomb gauge.  We verify that in the
same limit, the Gauss-BRST Ward identity holds, which is the functional
analog of the operator statement that a BRST transformation is generated by
the Gauss-BRST charge.  As a consequence, $gA_0$ is invariant under
renormalization, whereas in a covariant gauge, no component of the gluon
field has this property.
\end{abstract}

\end{titlepage}
\renewcommand{\thefootnote}{\arabic{footnote}}
\setcounter{footnote}{0}

\section{Introduction}

Although different gauges are formally equivalent, some are simpler than
others, or may have attractive properties.  Covariant gauges are well
adapted to perturbative expansion and renormalization.  However in QCD we are
interested in confinement and eventually in bound-state problems which are
inherently non-perturbative.  For such puposes, non-covariant gauges such as
the Coulomb gauge, the Weyl gauge, the axial gauge or the light-front gauge
may be attractive.  These gauges are considered ``physical'' in the sense that
the space of states is believed to be unitary and does not involve ghosts. 
(For a discussion of various gauges see \cite{physicalgauges}.)  Indeed
non-covariant gauges such as the Coulomb gauge and the light-front gauge have
recently been used to investigate confinement in QCD \cite{Z1, X}.   

	However it is a fact that at the level of quantum field theory, the well
established, renormalizable gauges for QCD are, on the one hand,
covariant, and on the other, involve ``unphysical'' particles.  These are
the fermi-ghosts that are needed to cancel the unphysical gluon degrees of
freedom.  One would like to know whether or not the physical gauges really
exist in the sense of perturbatively renormalizable quantum field theories,
and whether they are really unitary in the sense that they may be expressed
without ghosts, in terms of the two transverse degrees of freedom of the
gluon.  We shall see that for the Coulomb gauge, the answer to the first
question is ``yes'' and to the second, a slightly qualified ``no''.

	The point of view which we adopt in the present article is that the BRST
formulation provides a reliable method of quantizing and perturbatively
renormalizing non-Abelian gauge theories.  (For a review see \cite{Baulieu}
and \cite{Piguet&Sorella}.)  The existence and properties of physical or
canonical gauges will be investigated deductively starting from the BRST
formulation.  To be sure, this inverts the historical order in which gauge
theories were first canonically quantized, and subsequently the BRST method
was found; however the canonical method has remained heuristic, and to this
day does not allow systematic renormalization.

	There are two different problems raised by
the commonly used ``physical gauges'': (i) the breaking of Lorentz covariance
and (ii) an arbitrariness due to incomplete gauge fixing.  For example the
Coulomb gauge condition
$\grad \cdot \vec{A} = 0$ obviously breaks Lorentz invariance.  It is also
an incomplete gauge-fixing in the sense that it leaves a one-parameter
family of gauge transformations arbitrary, namely gauge transformations
$g(t)$ that are independent of the spatial coordinate
$\vec{x}$, but may depend on the time $t$.  Similarly, the Weyl gauge
condition $A_0 = 0$ leaves arbitrary a 3-parameter family of gauge
transformations, $g(\vec{x})$.  We call the dimension $\sigma$ of this
parameter space the ``degree of arbitrariness'' of the gauge, and we have
$\sigma = 1$ for the Coulomb gauge, and $\sigma = 3$ for the Weyl, the
axial, and the light-front gauges.  Not surprisingly, the degree of
arbitrariness of the gauge determines the dimension of the divergences of
Feynman integrals that are not controlled by usual ultraviolet regularization.

	Strictly speaking, incomplete gauge fixing with $\sigma > 0$ implies that the
correlation functions of charged fields actually vanish at generic space-time
separation.  For example in the Coulomb gauge, the arbitrariness under $g(t)$
implies that the correlation function of two charged fields vanishes at
unequal times,
$\langle \psi(\vec{x}, t) \psi^*(0, 0) \rangle = 0$ for $ t \neq 0$, 
even in abelian gauge theory.  This vanishing of correlation functions due to
gauge arbitrariness is not what one has in mind by a `physical' gauge, and it
is usually overcome in continuum gauge theory by additional gauge fixing by
more or less explicit prescriptions.\footnote{In lattice gauge theory,
gauge-fixing is frequently left incomplete.}  Incomplete gauge fixing would
appear to be the origin of ambiguities that occur in higher loop diagrams
\cite{Taylor}, and which make the formal Coulomb gauge, defined by canonical
quantization after elimination of the Coulomb-gauge constraints, not
particularly well-defined.  Consequently it is very misleading to speak of
{\em the} Coulomb gauge, as in the question, `What are the Feynman rules for
{\em the} Coulomb gauge?'.  Unless one is willing to accept the vanishing of
correlation functions of charged fields at unequal times, this question
cannot have a unique answer without further stipulation of the gauge
condition.  This applies to all gauges with $\sigma > 0$.

	We shall deal with both problems, Lorentz breaking and gauge arbitrariness,
by the device of an ``interpolating'' gauge.  For example the gauge condition
$-a\partial_0 A_0 + \grad \cdot \vec{A} = 0$, with 
$0 \leq a \leq 1$ interpolates between the Landau gauge, $ a = 1$, and the
Coulomb gauge, $a = 0$.  For $a > 0$ the gauge condition is regular, in the
sense that the degree of arbitrariness vanishes, 
$\sigma = 0$, but Lorentz invariance is broken for $a \neq 1$.  This allows
one to first address the problem of Lorentz breaking in a regular gauge,
and then to see if the singular limit $a \to 0$ yields finite correlation
functions.  In the present article we shall use and extended BRST symmetry
to control the violation of Lorentz invariance, and we shall then study the
Coulomb gauge limit of the Landau-Coulomb interpolating gauge.

		Use of an interpolating gauge and an extension of BRST symmetry to control
the violation of Lorentz invariance, was reviewed by Piguet \cite{Piguet},
particularly for the interpolating light-cone gauge.  Doust \cite{Doust}
used a gauge which interpolates between the Coulomb and Feynman gauge to
regularize the Coulomb gauge, and showed that extra terms in the Feynman
rules which he obtained in the Coulomb-gauge limit correspond
to an additional potential term obtained by Christ and Lee from an operator
ordering of their Coulomb Hamiltonian \cite{Christ&Lee}.  Difficulties with
renormalization in the Coulomb gauge were exhibited by Doust and Taylor
\cite{Doust&Taylor}.  The Weyl gauge ($A_0 = 0$) has
been studied by Rossi and Testa \cite{Rossi&Testa}, and by Cheng and Tsai
\cite{Cheng&Tsai}. 

        As commonly used in non-Abelian gauge theories,
BRST-invariance provides a substitute for invariance under local gauge
transformations which is broken by the gauge-fixing term.  In
Lorentz-covariant gauges, one uses the BRST method to enumerate the
independent divergent counter-terms necessary to ensure finitness of the
renormalized theory, while preserving all requirments of gauge invariance for
physical  quantities.  It is a powerful algebraic method of great
generality, relying as it does on the simplicity of invariance under a
generator $s$ that is nil-potent $s^2 = 0$.  

	In the first part of this article, we develop an extension of the BRST
method that also provides a substitute for
invariance under global Lorentz rotations when the gauge-fixing term
breaks global Lorentz invariance as well as local gauge invariance.  The
method is of considerable generality in that it does not rely on particular
properties of the symmetry which is broken by the gauge-fixing term, but only
that the symmetry operations form a Lie group, and it allows us to
explicitly enumerate all counter-terms.  
  
        For the class of interpolating gauges, defined by 
$(\alpha \partial)^\mu A_\mu = f$,
with $\alpha$ a
non-singular matrix, the partition function is formally given by the
Faddeev-Popov formula
\begin{equation}
 Z = \int dA \delta[(\alpha\partial)^\mu A_\mu - f]
        \det[(\alpha\partial)^\mu D_\mu(A)] \exp[ - S_{YM} ],
        \label{eq:FP}
\end{equation}
where $S_{YM}$ is the Euclidean Yang-Mills action.  Feynman graphs
contain denominators of the form $k_\mu \alpha^{\mu \nu} k_\nu$ and
$k^2$.  As long as $\alpha$ is non-singular, these
denominators provide the same degree of convergence in all
directions in $k$-space as the corresponding denominator $k^2$ in
covariant gauges.  Consequently in this class of interpolating
gauges, power counting of graphs is exactly the same as in
Lorentz-covariant gauges.  The problem of renormalizability is reduced to an
algebraic one of enumerating the form of possible local divergent terms, which
we control by extended BRST-invariance.  On the contrary, because of
gauge arbitrariness in the limiting cases of the Coulomb, light-cone or other
singular gauges, the degree of convergence depends on the direction in
k-space, and a more detailed analysis is required to determine if the limit
is finite.

	In the second part of the present article, we analyse the singular Coulomb
gauge limit, $a \to 0$, from the Landau-Coulomb
interpolating gauge.  For this purpose we express the
partition function $Z$ as a functional integral in phase space, and then
make a linear shift in the field variables in order to exhibit a
symmetry ($r$-symmetry) between the fermi and bose unphysical degrees of
freedom.  Individual closed fermi-ghost loops and closed
unphysical bose loops diverge like
$a^{-1/2}$, but they cancel pairwise by virtue of the $r$-symmetry. 
Consequently the correlation functions are finite in the limit $a \to 0$ from
the Landau-Coulomb gauge.  This remains true for the renormalized
correlation functions.  (See remark 1 at the end of sect.~9.)  However we also
find that there are one-loop graphs that vanish like
$a^{1/2}$, and that are missing in the formal ($a = 0$)
Coulomb gauge, but which cannot be neglected because they give a finite
contribution when inserted into the graphs that diverge like
$a^{-1/2}$.  It remains a logical possibility that these two-loop graphs,
that are missing in the formal Coulomb gauge, are mere gauge
artifacts that decouple from expectation values of all gauge-invariant
quantities such as a Wilson loop.  However there is at the moment no argument
to show that they do.  Indeed unless for some reason these two-loop graphs
decouple from all physical amplitudes, then the ghosts do not decouple in the
Coulomb gauge limit, and the Coulomb gauge is not unitary in the usual sense
of being a canonical theory of the transverse gluon degrees of freedom.  

	Nevertheless we find that correlation functions of the Coulomb-gauge
limit of the Landau-Coulomb interpolating gauge do exist, and moreover they
display a kind of simplicity that is absent from covariant gauges.  A certain
Gauss-BRST Ward identity holds in the Coulomb gauge limit which implies,
among other things, that the time-time componant of the gluon propagator
$g^2D_{00}$ is a renormalization-group invariant and thus depends only on a physical mass,
$\Lambda_{QCD}$, but not on the ultra-violet cut-off, $\Lambda$, nor the
renormalization mass, $\mu$, which may make it a useful order parameter for
color confinement.  No component of the propagator has this property
in a covariant gauge.

\section{Interpolating Gauges}

        In this section we introduce interpolating gauges for various
familiar classical gauges.  The Landau and Coulomb gauges are defined by
$-\partial_0A_0 + \grad \cdot \vec{A} = 0$ and $\grad \cdot \vec{A} = 0$.  The
Weyl and axial gauges are frequently defined by $A_0 = 0$ and $A_3 = 0$
respectively.  However if periodic boundary conditions are introduced in time
or space, the conditions $A_0 = 0$ and $A_3 = 0$ are too strong, and cannot be
maintained.  For they fix to unity the values of straight-line Wilson loops 
${\rm tr} P \exp(\int dx^\mu A_\mu)$ that close by periodicity, which however
are gauge-invariant objects.  We take instead as the Weyl and axial gauge
conditions the weaker conditions
$\partial_0A_0 = 0$ and $\partial_3A_3 = 0$.  In momentum space these read
$k_0\tilde{A}_0(k) = 0$ and $k_3\tilde{A}_3(k) = 0$, so the weaker
conditions differ from the stronger ones by zero modes only.  Similarly for
the light-front gauge condition, instead of $- A_0 + A_3 = 0$ we take
$(\partial_0 + \partial_3)(- A_0 + A_3) = 0$.  

	All these gauge conditions
have the linear form $(P\partial) \cdot A = 0$, where in the various cases
$P_\mu^\nu$ is the projector
\bea
{\rm Landau:} & P_\mu^\nu & = \delta_\mu^\nu = {\rm diag} (1, 1, 1,1) \CR
{\rm Coulomb:} &P_\mu^\nu & = {\rm diag} (0, 1, 1, 1)   \CR 
{\rm Weyl:} &P_\mu^\nu & = {\rm diag} (1, 0, 0, 0)   \CR 
{\rm axial:} &P_\mu^\nu & = {\rm diag} (0, 0, 0, 1)   \CR 
{\rm light-front:} &P_3^3& = P_3^0 = P_0^3 = P_0^0 = 1/2  
\ \ \ \ {\rm and} \ \  P_\mu^\nu = 0 \ \  {\rm otherwise}.
\eea
These projectors have a null space of dimension $\sigma =$ 0, 1, 3, 3, and 3
respectively, where $\sigma$ is the degree of arbitrariness of the gauge, as
defined in the Introduction.

 To separate the problem of violation of Lorentz invariance by the
gauge-fixing condition from the problem of the arbitrariness of the classical
gauges, we introduce an interpolating gauge defined by the condition
$(\alpha \partial) \cdot A = 0$.  Here $\alpha$ is the numerical
matrix
\be
\alpha \equiv P + a Q, 
\ee
where $P$ is one of the above projectors, $Q \equiv (1-P)$ is the orthogonal
projector, and $a$ is real, in the interval $0 \leq a \leq 1$.  These gauges
interpolate between the Landau gauge, at 
$a = 1$, and any one of the above singular classical gauges, which is achieved
at $a = 0$.  

	For the quantum field theory we consider the slightly more general
gauge condition $(\alpha \partial) \cdot A = f$.  By the usual
Faddeev-Popov argument, the partition function,  eq.~(\ref{eq:FP}),  is
expressed in terms of the local Faddeev-Popov action, 
\begin{equation}
        S_{\rm FP}(A, c, \bar{c}) \equiv S_{\rm YM}(A) 
+ \int d^4x \ \{ \ (2\beta)^{-1}[(\alpha \partial) \cdot A]^2 
+ (\alpha\partial) \bar{c} \  \cdot D(A) c \ \}   \label{eq:FPaction}
\end{equation} 
where $D(A)$ is the gauge-covariant derivative 
$[D_\mu(A)c]^a \equiv \partial_\mu c^a + f^{abd} {A_\mu}^b c^d$, and
$\beta$ is a gauge parameter.  

	From this action, one reads off the ghost propagator
\be
G = -i \ (k \cdot k')^{-1},
\ee
where
\be
k' \equiv \alpha k = Pk + aQk.
\ee
Similarly the gluon propagator $D$
is obtained from the quadratic part of the gluon action $(1/2)(A, KA)$ by 
$K^{\lambda\mu}D_{\mu\nu} = -i\delta_\nu^\lambda$.  From the Faddeev-Popov
action we have
\be
K^{\mu\nu} = k^2 g^{\mu\nu} - k^\mu k^\nu + \beta^{-1} k'^\mu k'^\nu,
\ee 
and one easily verifies that the gluon propagator is given by
\be
       D_{\mu\nu} = -i \ k^{-2} [ g_{\mu\nu} 
- (k \cdot k')^{-1} (k_\mu k'_\nu + k'_\mu k_\nu) 
+ (k \cdot k')^{-2}( \beta k^2 + k'^2) k_\mu k_\nu ] \ .
\ee
As long as $\alpha$ is a non-singular matrix, namely for $ a > 0$, convergence
of Feynman integrals is independent of direction in momentum space.  The
familiar power counting arguments hold, and Feynman integrals may be
regularized by dimensional regularization.

	We now consider some special cases.  A Landau-type interpolating gauge is
obtained at $\beta = 0$.  In this case the propagator satisfies the
generalized transversality condition $k'^\mu D_{\mu \nu} = 0$.  For $a=1$,
we have the Landau-gauge propagator, so this gauge interpolates smoothly
between the Landau gauge and the classical singular gauges.

	A Feynman-'tHooft type gauge is obtained by choosing $\beta$ so that
the double pole becomes a simple pole. For Coulomb, Weyl and axial gauges,
the projector
$P$ commutes with the metric tensor $g =$ diag( -1, 1, 1, 1), and we have
$(Pk)\cdot (Qk) = 0$.   In these gauges the double pole is eliminated by
setting $\beta = a$, for we have
\be
\beta k^2 +k'^2 = a [(Pk)^2 + (Qk)^2] + (Pk)^2 + a^2(Qk)^2
= (1+a) k \cdot k',
\ee
which gives the propagator
\be
       D_{\mu\nu} = -i \ k^{-2} \{ g_{\mu\nu} 
+ (k \cdot k')^{-1} [ - k_\mu k'_\nu - k'_\mu k_\nu + (1+a)k_\mu k_\nu ] \} 
\ .
\ee
This propagator has the attractive 'tHooft-type property
that it is block diagonal in the
$P$-$Q$ subspaces.  

	On the other hand, for the light-front gauge $(Pk) \cdot (Qk) \neq 0$, but
$Pk$ is a null vector, 
$(Pk)^2 = 0$. In this case the double pole is eliminated by setting
$\beta = a^2$, for we have
\bea
\beta k^2 +k'^2 = a^2[2(Pk)\cdot(Qk) + (Qk)^2] 
+ 2a(Pk)\cdot(Qk) + a^2(Qk)^2   \CR
= 2a k \cdot k',
\eea
which gives the propagator
\be
       D_{\mu\nu} = -i \ k^{-2} \{ g_{\mu\nu} 
+ (k \cdot k')^{-1} [ - k_\mu k'_\nu - k'_\mu k_\nu + 2ak_\mu k_\nu ] \} 
\ .
\ee
In the last two expressions for $D_{\mu\nu}$, the Feynman gauge is
obtained at $a = 1$, so these gauges interpolate smoothly between the
Feynman gauge and the classical singular gauges.  

	We write these expressions explicitly for interpolating
Coulomb gauges. In this case we have $k' = Pk + aQk = (ak_0, \vec{k})$, and
the ghost propagator is given by
\be
G= -i \ {{1\over {{{\vec k}^2} - ak_0^2}}} \ .
\ee
For the gauge which interpolates between the Landau and the Coulomb gauges,
the gluon propagator is given by
\be
i \ D_{ij} = {{1\over {k^2}}}
\left( g_{ij}-{{ k_i k_j}\over { {\vec k}^2 }} \right)
- {{ k_i k_j}\over { {\vec k}^2 }} { {a^2 k_0^2} \over 
{ ({\vec k}^2 - ak_0^2 })^2 }
\ee
\be
i \ D_{0i}= - { { a k_0 k_i }\over { ( {\vec k}^2 - ak_0^2 } )^2 }
\ee
\be
i \ D_{00}= - { { {\vec k}^2 }\over { ( {\vec k}^2 - ak_0^2 } )^2 } \ .
\ee
These expressions are obtained by partial fractionation, and there is no
singularity at $\vec{k} = 0$ for $a > 0$.

 It is easy to understand intuitively how the Coulomb-gauge limit from the
Landau-Coulomb gauge fixes the gauge arbitrariness of the Coulomb gauge
discussed in the Introduction.  Under the residual gauge freedom of the
Coulomb gauge, $A_0$ transforms according to 
$A_0 \to g^{\dag}(t)A_0g(t) + g^{\dag}(t)\partial_0g(t)$, where
the inhomogeneous term is $\vec{x}$-independent.  With periodic boundary
conditions, the Landau-Coulomb gauge condition 
$a\partial_0A_0 = \vec{\nabla} \cdot \vec{A}$ for $a > 0$ gives
$\partial_0 \int d^3x A_0 = 0$.  However, as one sees from the above
expression for the $A_0-A_0$ propagator, $D_{00}$ vanishes at $\vec{k} = 0$
for all finite $a$, so the stronger condition
$\int d^3x A_0 = 0$ in fact holds in the Landau-Coulomb gauge for all finite
$a$.  This provides the additional gauge-fixing condition needed to make
the limit $a \to 0$ well defined.  By contrast $D_{00}$ in the
Feynman-Coulomb gauge, given below, becomes ill defined at $\vec{k} = 0$
with periodic boundary conditions, in the limit $a \to 0$.  

	For the gauge which interpolates between the Feynman and the
Coulomb gauges one has
\be
i \ D_{ij} = {{1\over {k^2}}}
\left( g_{ij}-{{ k_i k_j}\over { {\vec k}^2 }} \right)
+ {{ k_i k_j}\over { {\vec k}^2 }} { {a} \over { {\vec k}^2 - ak_0^2 } }
\ee
\be
i \ D_{0i}= 0
\ee
\be
i \ D_{00}= - {{1\over {{{\vec k}^2} - ak_0^2}}} \ .
\ee
There is no mixing of space and time components of the gluon propagator in
this gauge.

	These expressions for the propagators are quite
illuminating. The transverse part of $D_{ij}$ is the
Coulomb gauge propagator.  The parameter $a$ acts as a
regulator for simultaneity in the Coulomb gauge.  These expressions
imply   exact compensations
 between the ``unphysical'' contributions in internal loops between
gluon and ghost propagators.
 The main
thing is of course that these compensations occur because the poles of
the propagators of the unphysical fields with opposite statistics 
sit at the same point, i.e, at $ {\vec k}^2 - ak_0^2 =0$.

 For completeness, we indicate the form of propagators in the 
interpolating gauges for the light-front gauge quoted in \cite{Grange},
with 
$k_\mu' = (\alpha_{\rm LFG}k)_\mu$ given by 
\be 
k_1'= ak_1, \ \ k_2' =  ak_2, \ \ \ 
k_3' = {{(1+a)\over 2}}k_3 + {{(1-a)\over 2}}k_0, \ \ \ 
k_0' = {{(1-a)\over 2}}k_3 + {{(1+a)\over 2}}k_0.
\ee
We have $k \cdot k' = a(k_1^2 + k_2^2) + {{(1+a)\over 2}}(k_3^2 - k_0^2)$.
For the gauge which interpolates between the light-front and Feynman gauges,
namely, with $\beta = a^2$ which eliminates the double pole in the gluon
propagator, the above expression for the gluon propagator reads
\bea
D_{ij} & = & -i \ {{\delta_{ij}\over {k^2}}}    \CR
D_{i-} & = & 0     \CR
D_{i+} & = & i \ (1-a) \ {{k_i(k_3 + k_0)\over {k^2 \ k\cdot k'}}}   \CR
D_{--} & = & 0  \CR
D_{+-} & = & {{-2ia \over {k \cdot k'}}}    \CR
D_{++} & = & 2 \ i \ (1-a) \  
  {{(k_3 + k_0)^2 \over {k^2 \ k \cdot k'}}}, 
\eea 
where $i, j = 1, 2$, $D_{\mu\pm} = D_{\mu3} \pm D_{\mu0}$, and
$D_{\pm \pm} = D_{3\pm} \pm D_{0\pm}$.  We observe that 
$D_{\mu -} = 0$ at $a = 0$, and the light-front gauge condition is satisfied.

\section{BRST symmetry for local gauge and global Lorentz invariance}
Suppose that we have a Lie algebra with basis $X_i$ and structure
constants $f_{ijk}$, so $[X_i,X_j] = f_{ijk}X_k$.  According to the
BRST method, for each generator $X_i$ we introduce a corresponding Grassmann
or ghost variable $c_i$.  The BRST operator $s$ acts on these variables
according to
\begin{equation}
sC_i = -  {1\over 2} f_{ijk} C_j C_k.
\end{equation}
It is nilpotent, $s^2 = 0$.  The preceding relation is isomorphic to the
action of Cartan's exterior differential operator d acting on the
Maurer-Cartan form $\omega = \omega_it^i = g^{-1}dg$ of the Lie group. 

                We wish to apply this method to the Lie group which
consists of local
gauge transformations and global Lorentz transformations.  The structure
constants of this group are given by
\begin{eqnarray}
& [ G^a(x), G^b(y) ] = f^{abc}\delta(x-y)G^c(x)  \nonumber  \CR
& [ H_{\mu\nu}, G^a(x) ] 
        = - (x_\mu \partial_\nu - x_\nu \partial_\mu ) G^a(x)  \nonumber   \CR 
& [H_{\kappa \lambda}, H_{\mu \nu}] 
        = g_{\lambda \mu} H_{\kappa \nu} 
        - g_{\kappa\mu} H_{\lambda \nu}
        - g_{\lambda \nu} H_{\kappa \mu}
        + g_{\kappa \nu} H_{\lambda \mu}.
\end{eqnarray}

        According to the method described above, corresponding to the
local generators $G^a(x)$ we introduce the usual anti-commuting
Grassmann field variables $c^a(x)$ and corresponding to the $H_{\mu\nu}$ 
and we introduce the global generators $V_{\mu\nu} = - V_{\nu\mu}$, so $C_i
= (c^a(x), V_{\mu \nu})$.  In 4-dimensional space-time there are 6
independent generators $V_{\mu\nu}$.  For the structure constants just
found, the BRST operator
$s$ acts according to
\begin{eqnarray}
 sc^a(x) & = &- {1\over 2} f^{abc} c^b(x)c^c(x)  
   + \: z \:{V_\mu}^\nu x^\mu \partial_\nu c^a(x)    \label{eq:sc}   \\   
 s{V_\lambda}^\nu & = & - {V_\lambda}^\mu {V_\mu}^\nu,       \label{eq:sV}
\end{eqnarray}
where the parameter $z$ will be determined shortly.  Because $V_{\mu \nu}$
is a Grassmann variable, $(V^2)_{\kappa \nu} = V_{\kappa \lambda}
g^{\lambda \mu} V_{\mu \nu}$  is an anti-symmetric matrix,
$(V^2)_{\mu \nu} = - (V^2)_{\nu \mu}$.  Equation \ref{eq:sc} determines the
normalization of the ghost field
$c^a(x)$, and eq. \ref{eq:sV} determines the normalization of Grassmann
variables $V_{\mu,\nu}$.  The parameter $z$ is
most easily determined by requiring that $s$ be nil-potent, $s^2 = 0$,
which gives $z = -1$.  We could as easily have derived the corresponding
result for the Poincare group.\footnote{ Use of a ``large'' BRST
operator in the present context was suggested to us by Massimo Porrati}

        The BRST operator associated to the Lie algebra just defined is
of the form $s = s_g + s_L$, where $s_g$ and $s_L$ satisfy 
$(s_g)^2 = (s_L)^2 = s_gs_L + s_Ls_g = 0$.  On the fields $c^a(x)$ and
${V_\mu}^\nu$ they act according to
\begin{eqnarray}
        & s_gc^a(x) = - {1\over 2} f^{abc} c^b(x)c^c(x)    \nonumber     \\
        & s_Lc^a(x) =  {V_\mu}^\nu x^\mu \partial_\nu c^a(x)   \nonumber     \\
 & s_g{V_\lambda}^\nu = 0    \nonumber      \\
        & s_L{V_\lambda}^\nu = - {(V^2)_\lambda}^\nu.      \label{eq:MC}
\end{eqnarray}
The BRST oprator
$s$ defined here may be viewed as a ``large'' BRST operator, which is the
usual BRST operator $s_g$ for the local gauge group extended by the BRST
operator $s_L$ for the global Lorentz group.\footnote{ The Lorentz
rotations are a subset of general reparametrization.  If we define the
vector $\xi^\mu= V^\mu_\nu x^\nu$, out of the constant ghosts $V$ and
the coordinates $x^\mu$, we have $s \xi^\mu = \xi^\rho \partial_\rho
\xi^\mu$ and $s_Lc=\xi^\mu\partial_\mu c$.}

        To determine the action of the BRST operator on the connection
${A_\mu}^a(x)$, we could start with the familiar transformation law of the
connection under local gauge and global Lorentz transformation,
\begin{eqnarray}
        & G(\omega){A_\mu}^a(x) = \partial_\mu \omega^a(x) 
        + f^{abc}{A_\mu}^b(x) \omega^c(x)   \nonumber     \\
 & H(\epsilon){A_\mu}^a(x) = {\epsilon_\kappa}^\lambda 
        x_\kappa \partial_\lambda {A_\mu}^a(x)
        + {\epsilon_\mu}^\nu {A_\nu}^a(x).
\end{eqnarray}
A more economical way is to construct the most general expression 
$s{A_\mu}^a = (s_g + s_L){A_\mu}^a$ which satisfies $s^2=0$.  Suppose that
$s_L$ acts according to
\begin {equation}
        s_L{A_\mu}^a(x) = z_1 {V_\kappa}^\lambda x^\kappa \partial_\lambda  
        {A_\mu}^a(x) + z_2 {V_\mu}^\nu {A_\nu}^a(x).   \label{eq:Lorentz}
\end{equation}
Here $z_1$ and $z_2$ are parameters that are determined by the condition
$(s_L)^2 = 0$.  From eq.~(\ref{eq:MC}) one obtains $z_1 + (z_1)^2 = 0$ and
$z_2 + (z_2)^2 = 0$.  We take $z_1 = -1$ because $z_1 = 0$  gives a trivial
field transformation law.  The components of $A_\mu$ transform either like
scalars ($z_2 = 0$) or a vector ($z_2 = -1$).  We take the vector case and
obtain
\begin {equation}
        s_L{A_\mu}^a(x) = - {V_\kappa}^\lambda x^\kappa \partial_\lambda  
        {A_\mu}^a(x) - {V_\mu}^\nu {A_\nu}^a(x).
\end{equation}

	Finally, suppose that $s_g$ acts on $A$ according to
\begin{equation}
        s_g{A_\mu}^a = {z_{1\mu}}^\nu \partial_\nu c^a 
        + {z_{2,\mu}}^\nu f^{abc}{A_\nu}^b c^c.
\end{equation}
We obtain  from $(s_g)^2 = 0$ that  ${z_{2,\mu}}^\nu = {\delta_\mu}^\nu$, and
${z_{1,\mu}}^\nu$ remains arbitrary. The condition $s_gs_L + s_Ls_g = 0$
gives $z_1V = Vz_1$.  Because ${V_\mu}^\nu$ is arbitrary, $z_1$ is of the
form ${z_{1\mu}}^\nu = z {\delta_\mu}^\nu$.  We write  
${A_\mu}^a \equiv z {A'_\mu}^a$, and obtain for the BRST
operator $s = s_g + s_L$,
\begin{equation}
        s{A_\mu}^a = \partial_\mu c^a + f^{abc}{A_\mu}^b c^c
        - {V_\kappa}^\lambda x^\kappa \partial_\lambda  
        {A_\mu}^a(x) - {V_\mu}^\nu{A_\nu}^a(x),        \label{eq:sA}
\end{equation}
where we have dropped the prime on $A'$.  This completes the determination
of the action of the BRST operator on the basic fields $A$ and $C$.

\section{Extended action}

        The partition function in  eq.~(\ref{eq:FP})  may be expressed in terms of
the local Faddeev-Popov action, 
\begin{equation}
        S_{\rm FP}(\Phi) \equiv S_{\rm YM}(A) + \int d^4x 
[-(\alpha \partial)^\mu b A_\mu + (\alpha \partial)^\mu \bar{c}D(A)_\mu c
 +{{\beta \over 2}}
   b^2 ],     \label{eq:FPaction}
\end{equation}
where $D(A)$ is the gauge-covariant derivative 
$[D_\mu(A)c]^a \equiv \partial_\mu c^a + f^{abc} {A_\mu}^b c^c$, and
$\Phi$ represents the set of fields $\Phi = (A, c, \bar{c}, b)$.  We
introduce a corresponding set of sources, 
$J = (J_A,J_c, J_{\bar{c}}, J_b)$, and write 
\begin{equation}
        Z(J) = \int d\Phi \exp[ - S_{\rm FP}(\Phi) + (\Phi, J)],    
\label{eq:partition}
\end{equation}
where $d\Phi \equiv dA dc d\bar{c} db$, and 
\begin{equation}
(\Phi, J) \equiv \int d^4x 
(A \cdot J_A + c \cdot J_c + \bar{c} \cdot J_{\bar{c}} + b \cdot J_b).
\end{equation}

        The Faddeev-Popov action is not invariant under Lorentz
transformations because of the appearance of the numerical matrix
$\alpha^{\mu \nu}$.  Consider instead the extended action
\begin{equation}
        S_{\rm ext}(\Phi, V) \equiv S_{\rm YM}(A) 
        - s\int d^4x [ (\alpha \partial)^\mu \bar{c} A_\mu -
{\beta \over 2}
\bar{c}b ],
 \label{eq:extendedaction} \end{equation}
where $s$ is the ``large'' BRST operator that expresses the
substitute gauge and Lorentz transformations.  Its action on $A$, $c$ and
$V$ is defined in eqs.~(\ref{eq:sA}), (\ref{eq:sc}) and (\ref{eq:sV}), and
its action on $\bar{c}$ and $b$ is defined by $s\bar{c} = b$ and $sb = 0$,
which preserves $s^2 = 0$.  Because the Yang-Mills action $S_{\rm YM}(A)$
is both gauge and Lorentz invariant, it is invariant under the ``large''
BRST operator $sS_{\rm YM}(A) = 0$, and consequently so is the extended
action, 
$sS_{\rm ext}(\Phi, V) = 0$.   The extended action differs from
the Faddeev-Popov action by terms linear in the
global Grassmann variables $V$ introduced in the preceding section,
\begin{equation}
        S_{\rm ext}(\Phi, V) = S_{\rm FP}(\Phi) 
- \int d^4x (\alpha \partial)^\mu \bar{c} 
        ({V_\kappa}^\lambda x^\kappa \partial_\lambda A_\mu +{V_\mu}^\nu A_\nu).
\end{equation}
We treat the variables $V_{\mu \nu}$ as external
sources,\footnote{Equivalently we may treat the vector 
$\xi_\mu = V_\nu^\mu x^\nu$ as an extended source.} and define the
extended partition function
\begin{equation}
        Z(J, V) \equiv \int d\Phi 
        \exp[ - S_{\rm ext}(\Phi, V) + (\Phi, J)].
\end{equation}
The original partition function is obtained from it by $Z(J) = Z(J, 0)$. 
Because there are 6 independent global Grassmann variables $V$,
there are, in all, $2^6$ terms in the expansion of $Z(J, V)$ in powers of
$V$.  They are related by the symmetry generated by the large $s$. 

	The usual argument that the expectation values of gauge-invariant
observables are independent of the gauge parameters must be slightly modified
because the variable $V$ is not integrated over.  We consider only
$s$-invariant observables $W$ are indepndent of $V$.  We shall show that
$\langle W \rangle$ is
independent of the $\alpha$ matrix when the external source $V$ is set to 0. 
We also set all
sources $J$ to 0, and we have
\bea
\partial\langle W \rangle / \partial \alpha^{\mu\nu}
= \int d\Phi \  W  \int d^4x \ s(\partial_\nu \bar{c} \ A_\mu) 
\exp(-S_{\rm ext}) |_{V = 0} \CR
= \int d^4x \ \int d\Phi \ s[W \ \partial_\nu \bar{c} \ A_\mu
\exp(-S_{\rm ext}) \ ] |_{V = 0},
\eea
where we have used $sW = sS_{\rm ext} = 0$.  At $V = 0$ we have $s = s_g$,
where $s_g$ is a derivative with respect to the variables of integration
$\Phi = (A, c, \bar{c}, b)$.  This gives 
$\partial\langle W \rangle / \partial \alpha^{\mu\nu} = 0$, as asserted. 
We conclude that for physical observables, the interpolating
gauges gives the same expectation values as the covariant gauges.  In
particular they are independent of the gauge parameter $\alpha$, and
similarly for $\beta$.

\section{Quantum Effective Action}

        To exploit BRST symmetry in renormalization theory, it is helpful
to also
introduce sources for the BRST transforms that are non-linear in the
fields.  We therefore define the (fully extended) action
\begin{eqnarray}
 \Sigma(\Phi, V, K, L, M) 
& \equiv & S_{\rm ext} + (K, sA) + (L, sc) + M \cdot sV,    \CR
& = & S_{\rm ext} + s[ - (K, A) + (L, c) + M \cdot V \ ]
\label{eq:fea}
\end{eqnarray}
where ${K_\mu}^a(x)$ and $L^a(x)$ are the usual sources for
$s{A_\mu}^a(x)$ and $sc^a(x)$, and we have introduced a corresponding
source $M^{\mu \nu} = - M^{\nu \mu}$ for $sV = - V^2$, with $M \cdot sV
= {1\over 2}M^{\mu \nu} sV_{\mu \nu}$.  These sources are not acted on by $s$,
$sK = sL = sM = 0$.  The action $\Sigma$ is invariant under the ``large''
BRST operator,$s \Sigma =0$.  

        We define the corresponding partition function
\begin{equation}
Z(J, V, K, L, M) \equiv \int d\Phi \exp[ - \Sigma + (\Phi, J) ].
\end{equation}
It satisfies
\begin{equation}
{{
\delta Z }
\over
{
\delta
     M^{\mu \nu}  }}
 = - (V^2)_{\mu \nu} Z.
\end{equation}  

        The BRST operator $s$ has been defined as a linear differential operator
that acts on (and mixes) the variables $\Phi = (A, c,\bar{c}, b)$ and $V$. 
Because only the $\Phi$ variables are integrated over, it is convenient to
decompose $s$ according to $s = s_{\Phi} + s_V$, where $s_{\Phi}$ acts
only on the $\Phi$ variables, and $s_V$ only on $V$, so 
$s_{\Phi} V = s_V \Phi = 0$.  The explicit form of $s_V$ is
\begin{equation}
        s_V \equiv (sV) \cdot
{{
 \delta} \over 
 { \delta V }}
= - (V^2)_{\mu\nu}
{{
 \delta}\over {
   \delta V_{\mu\nu}}}
\end{equation}

  By the invariance of $\Sigma$ with
respect to $s = s_{\Phi} + s_V$, we have
\begin{equation}
(sV) \cdot
{{\delta Z 
}\over{ \delta V }}
= [ - (J_A,
{{
 \delta
}\over{
   \delta K}}
) - (J_c,
{{
 \delta
}\over{
   \delta L}}
) 
- (J_{\bar{c}},
{{
 \delta
}\over{
   \delta J_b
 }} 
) \ ]Z
\end{equation}

        The free energy $W(J, V, K, L, M) \equiv \ln Z(J, V, K, L, M)$,
satisfies
the corresponding equations
\begin{equation} 
{{\delta W }\over {\delta M^{\mu\nu}}} = -sV_{\mu \nu} 
 \label{eq:W1}
\end{equation}
and
\begin{equation}
 (J_A, {{\delta W}\over {\delta K}})
                       + (J_c, {{\delta W}\over {\delta L}})
+ (J_{\bar{c}},
  {{\delta W}\over {\delta J_b }})
+ (sV) \cdot
  {{\delta W}\over {\delta V   }}
 = 0.    \label{eq:W2}
\end{equation}

        We make a Legendre transformation from the variables
$J = (J_A,J_c, J_{\bar{c}}, J_b)$, and the free energy $W$ to the
external field variables $\Phi = (A, c, \bar{c}, b)$, and the quantum
effective action $\Gamma$,
\begin{equation}
\Gamma(\Phi, V, K, L, M) = (\Phi, J) - W(J, V, K, L, M)
\end{equation}
where
\begin{eqnarray}
 A_\mu ={{\delta W
 }\over {\delta  J_{A_\mu} }}
\quad
 c     ={{\delta W
 }\over {\delta  J_c       }}
\quad
\bar c ={{\delta W
 }\over {\delta  J_{\bar c
 } }}
\quad
 b     ={{\delta W
 }\over {\delta  J_b       }}
  \nonumber  \\
\end{eqnarray}

\begin{eqnarray}
 J_{A_\mu} =
{{\delta \Gamma }\over {\delta A_\mu}}\quad
  J_{c}     =
{{\delta \Gamma }\over {\delta c}}
\quad
 J_{\bar c} =
{{\delta \Gamma }\over {\delta \bar c
}}
 \quad
 J_{b}     =
{{\delta \Gamma }\over {\delta b}}
  \nonumber  \\
\end{eqnarray}
and
\begin{equation}
{{\delta \Gamma }\over {\delta V}}
=-
{{\delta W }\over {\delta V}}
; \;\;\;\; 
{{\delta \Gamma }\over {\delta K}}
=-
{{\delta W }\over {\delta K}}
; \;\;\;\; 
{{\delta \Gamma }\over {\delta L}}
=-
{{\delta W }\over {\delta L}}
; \;\;\;\; 
 {{\delta \Gamma }\over {\delta M}}
=-
{{\delta W }\over {\delta M}}
; \;\;\;\; \\
\end{equation}
Here and elsewhere, all derivatives with respect to fermionic variables
are left derivatives.  In terms of $\Gamma$, eqs. (\ref{eq:W1}) and
(\ref{eq:W2}) give
\begin{equation}
{{
\delta \Gamma
 }\over {
 \delta M^{\mu \nu} }}
= sV_{\mu \nu} = - (V^2)_{\mu \nu}.
\end{equation}
and
\begin{equation}
 (
 {{
\delta \Gamma
 }\over {
 \delta A           }}
,
   {{
\delta \Gamma
 }\over {
 \delta K           }}
)
+
( {{
\delta \Gamma
 }\over {
 \delta c           }}
,
   {{
\delta \Gamma
 }\over {
 \delta L           }}
)
+
 (
 {{
\delta \Gamma
 }\over {
 \delta \bar c      }}
,
  b
)
+
 {{
\delta \Gamma
 }\over {
 \delta V           }}
.
   {{
\delta \Gamma
 }\over {
 \delta M           }}
=0 \ .
\label{eq:Z-J1}
\end{equation} 
This type of equation, which was introduced in \cite{zinn}, now includes a
$V$-$M$ term.  Here is it assumed that there is no gauge
anomaly.  No Lorentz anomaly can occur in $D = 4$ dimensions.

        Because the gauge condition is linear, we may solve the equations of
motion to obtain the dependence of $\Gamma$ on the Lagrange multiplier
fields $\bar{c}$ and $b$.  As this is standard, we simply give the result
\cite{I&Z},
\begin{equation}
\Gamma(A, K, c, L, \bar{c}, b, V, M) = 
\int d^4x [ - (\alpha \partial)^\mu b A_\mu + ({\beta\over 2}) b^2 ]
+ \tilde{\Gamma}(A, K + (\alpha \partial) \bar{c}, c, L, V, M)\CR
\end{equation}
The property that the
$K$ and $\bar c$ dependences are only through the
 combination $K_\mu + \alpha\partial _\mu \bar c$
 can be imposed as a Ward identity in the class of linear gauges that we
consider.  This  plays an important role in the renormalisation program.

		The master  equation satisfied by 
$\tilde{\Gamma}(A, K, c, L, V, M)$ is symmetric in the pair $V, M$ and the
other variables,
\begin{equation}\label{master}
(
{{\delta \tilde \Gamma} \over {\delta A}},
{{\delta \tilde \Gamma} \over {\delta K}} )
+
(
{{\delta \tilde \Gamma} \over {\delta c}},
{{\delta \tilde \Gamma} \over {\delta L}} )
+
{{\delta \tilde \Gamma} \over {\delta V}}
\cdot
{{\delta \tilde \Gamma} \over {\delta M}} =0
.   
\end{equation}
$\tilde{\Gamma}$ has the simple dependence on $M$ given by
\begin{equation} {{
\partial
 \tilde{\Gamma} }\over
{ \partial M^{\mu \nu}  }}
= sV_{\mu \nu} 
= - (V^2)_{\mu \nu}.       \label{mmm}
\end{equation}

\section{Form of Divergences}

       The new $V-M$ term has the same structure as the other terms, so we
may use familiar arguments, which we now sketch, to determine the form of
possible divergences to each order in $\hbar$, when using a regulator that
preserves Lorentz and gauge symmetries.  We make a loop or
$\hbar$ expansion of $\tilde{\Gamma}$,  using any suitable regularization for
divergences.
\begin{equation}
\tilde{\Gamma} = \sum_n \tilde{\Gamma}^n .
\end{equation}
To find $\tilde{\Gamma}^0$, we observe that $\Sigma$, eq.~(\ref{eq:fea}), is
of the form
\begin{equation}
\Sigma(A, K, c, L, \bar{c}, b, V) = 
\int d^4x [ - (\alpha \partial)^\mu b A_\mu +  {1\over 2}  \beta b^2 ]
+ \tilde{\Sigma}(A, K + (\alpha \partial) \bar{c}, c, L, V, M)\CR   
\end{equation}
where
\begin{equation}
\tilde{\Sigma}(A, K, c, L, V) \equiv S_{\rm YM} + (K, sA) + (L, sc) 
+ M \cdot sV.
\end{equation}
This gives
\begin{equation}
\tilde{\Gamma}^0 = \tilde{\Sigma} .  
\end{equation}

	We will impose
 that $\tilde\Gamma$
 is renormalized while satisfying the master equation (\ref{master}). 
        From the $s$-invariance of $S_{\rm YM}$, and from 
$sA = \delta\tilde{\Sigma} / \delta K
$, 
$sc = \delta\tilde{\Sigma} / \delta L
$, and
$sV = \delta \tilde{\Sigma} / \delta M$, we have
\begin{equation}
 (
{{\delta \tilde \Sigma} \over {\delta A}},
{{\delta \tilde \Sigma} \over {\delta K}} )
+
(
{{\delta \tilde \Sigma} \over {\delta c}},
{{\delta \tilde \Sigma} \over {\delta L}} )
+
{{\delta \tilde \Sigma} \over {\delta V}}
\cdot
{{\delta \tilde \Sigma} \over {\delta M}}=0
\end{equation}
so the master equation (\ref{master}) is satisfied by 
$\tilde{\Gamma}^0 = \tilde{\Sigma}$.  We define the star product
\begin{equation}\tilde
\Gamma_a \ast \tilde
 \Gamma_b \equiv
(
{{\delta \tilde \Gamma_a
} \over {\delta A}},
{{\delta \tilde \Gamma_b
} \over {\delta K}} )
+
(
{{\delta \tilde \Gamma_a
} \over {\delta c}},
{{\delta \tilde \Gamma_b
} \over {\delta L}} )
+
{{\delta \tilde \Gamma_a
} \over {\delta V}}
\cdot
{{\delta \tilde \Gamma_b
} \over {\delta M}}
\\
\end{equation} 
To each order $n$ in $\hbar$, eq. (\ref{master}) reads
\begin{equation}
\sum_{p+q = n} \tilde{\Gamma}^p \ast \tilde{\Gamma}^q = 0. 
\label{eq:ordern}
\end{equation}

 We assume that renormalization has been done to order $n-1$, so that
$\tilde{\Gamma}^p$ for $p = 0, \cdots (n-1)$ is finite, and
that eq.~(\ref{eq:ordern}) is satisfied to order $n-1$.  We separate
the regular and  divergent parts of order $n$,
\begin{equation}
\tilde{\Gamma}^n =   \tilde{\Gamma}_R^n + \tilde{\Gamma}_{\rm div}^n,
\end{equation}
where the first term is the renormalized part of the n-th order effective
action and is finite.  By hypothesis, the only divergence in
eq.\ (\ref{eq:ordern}) comes from $\tilde{\Gamma}_{\rm div}^n$.  The divergent
part must satisfy eq.\ (\ref{eq:ordern}) separately, namely
\begin{equation}
\sigma \tilde{\Gamma}_{\rm div}^n = 0,  \label{eq:cohomology}
\end{equation}
where the linear operator $\sigma$, defined by
\begin{equation}
\sigma\Gamma \equiv \tilde{\Sigma} \ast \Gamma 
+ \Gamma \ast \tilde{\Sigma},
\end{equation}
has the explicit expression
\begin{eqnarray}
\sigma
 = \int d^4x \left(
{{
 \delta \tilde{\Sigma} } \over {
\delta K }}
{{
\delta }\over {\delta A}}
+
{{
 \delta \tilde{\Sigma} } \over {
\delta A }}
{{
\delta }\over {\delta K}}
+
{{
 \delta \tilde{\Sigma} } \over {
\delta L }}
{{
\delta }\over {\delta c}}
+
{{
 \delta \tilde{\Sigma} } \over {
\delta c }}
{{
\delta }\over {\delta L}}
+
{{
 \delta \tilde{\Sigma} } \over {
\delta M }}
{{
\delta }\over {\delta V}}
+
{{
 \delta \tilde{\Sigma} } \over {
\delta V }}
{{
\delta }\over {\delta M}}\right)
\CR
\end{eqnarray}
It is nilpotent $\sigma^2 = 0$.  Here $\sigma$ represents the
symmetry of the ``large'' BRST operator,  with the obvious decomposition
into local gauge and global Lorentz parts, $\sigma = \sigma_g + \sigma_L$,
that corresponds to $s = s_g + s_L$.  From eq. (\ref{mmm}) we have
\begin{equation}{{
\delta
 \tilde{\Gamma}^n }\over {
  \delta M }}
 = 0;   \;\;\; n \geq 1,
\end{equation}
so $\tilde{\Gamma}^n$ is independent of $M$.

        Consistent with the last equation, with locality of divergent
terms, with
global color invariance, with the ghost quantum numbers $(0, -1, 1, -2, 1)$
and dimensions $(1, 2, 1, 2, 1)$ of the variables $(A_\mu^a, K^{a, \mu},
c^a, L^a, V_{\mu \nu})$ on which $\tilde{\Gamma}^n$ depends,
eq.\ (\ref{eq:cohomology}) has the solution
\begin{equation}
\tilde{\Gamma}_{\rm div}^n = \int d^4x\left [ \ c_1 {1\over 4}
F_{\mu \nu}^2
        + \sigma
\left( K^{a \mu} {c_{2, \mu}}^\nu A_\nu^a + c_3 L^a C^a
\right )\ 
\right ],   \label{eq:solution}
\end{equation}
where $c_1$, ${c_{2, \mu}}^\nu$ and $c_3$ are divergent constants of order
$\hbar^n$.  The operator 
$x_\mu \partial_\nu - x_\nu \partial_\mu$ may appear in $\tilde{\Gamma}^n$ in
the combination $V^{\mu \nu} x_\mu \partial_\nu$.  However 
$x_\mu \partial_\nu - x_\nu \partial_\mu$ is dimensionless and carries no
ghost or global color quantum number so it does not affect our counting
arguments, which exclude the explicit appearance of $V$ in the last
equation.  However a $V$ dependence is introduced into $\tilde{\Gamma}^n$
from the definition of $\sigma$, so that $V$ appears in the expansion of the
$\sigma$-exact term.

        With this result we have achieved our goal of limiting
the number of possible divergences, by maintaining invariance under the
larger group of substitute gauge and Lorentz invariance.  Indeed only the
combination $\int d^4x F_{\mu \nu}^2$ is invariant under $\sigma$ without
being exact, of the form $\sigma X$.  ($\int d^4x F_{\mu \nu}^2$ is said to
be the cohomology of the operator $\sigma$.)   For if only invariance under
$s_g$ or
$\sigma_g$ were enforced, then the most general cohomology would be  
$\int d^4x (c_E E^2 + c_B B^2)$, if
ordinary rotational invariance is preserved by the gauge fixing, where $c_E$
and $c_B$ are {\em independent} renormalization constants.  Indeed, $E^2$ and
$B^2$ are separately invariant under $s_g$, and in \cite{Z1}, it was
necessary to {\em assume} $c_E = c_B$.  This is now established for the
gauges considered here. On the other hand the breaking of Lorentz invariance
by the gauge fixing does lead to the Lorentz non-invariant divergent terms 
$\sigma K^{a, \mu} {c_{2 \mu}}^\nu A_{\mu}^a$, which however are
exact $\sigma$-forms.  

        If ordinary rotational invariance is maintained by
the gauge fixing, then ${c_{2,\mu}}^\nu$ is a diagonal tensor with 
${c_{2,1}}^1 = {c_{2, 2}}^2 = {c_{2, 3}}^3 \neq {c_{2, 4}}^4$.  In
Lorentz-type gauges, defined by $\beta = 0$ in eqs. (\ref{eq:FPaction}) or
(\ref{eq:extendedaction}), the (possibly) divergent constant $c_3$
vanishes, $c_3 = 0$, by virtue of the factorization of the external ghost
momentum, as it does in the Landau gauge\cite{I&Zghost}.

\section{Multiplicative Renormalization}

\def\I{ \Sigma_\r(\Phi,K_\Phi, V)}

In previous sections we had implicitly absorbed the coupling constant $g$
into the SU(N) structure constant $f^{abc}$.  Since we are interested in the
perturbative expansion we now make the coupling constant explicit by the
substitution $f^{abc} \rightarrow gf^{abc}$.  

	We define
$\I_r$ as  the {\it local} part of $\Gamma=\Gamma_R +\Gamma_{{\rm div}}$. We
call $\I$ the renormalized action  since by inserting ${\rm exp}
 \I$ 
 in the path integral over  the $\Phi$,  all
 relevant local divergent  counterterms are present to determine finite Green
 functions of the fields $A,c,\bar c, b$ and of their BRST transformations
 which satisfy the BRST  master equations.

	The result found in  eq.~(\ref{eq:solution}), proves
 that the renormalized action   $\I_r$   has the following form:
\bea\label{action}
\I_r & = & \int d^4 x
\{ \ 
{1\over 4 }
| \ 
 F_
{\mu\nu}(Z^{\nu}_{A\mu} A_\nu, Z_g g)\ |^2
\CR
& + & Z_c
[ \ 
 (\alpha \partial)^\mu\bar{c} +K^\mu ]{Z _{A\mu}^{-1\nu}}
(\ 
\partial_\nu c+Z_g g  [\  Z_{A\mu}^{\rho} A_\rho\ ,\  c ]\ \big)
\CR
& - & {1\over 2}Z_c  Z_g g [\ c,\ c\ ]\ L 
+ {\beta\over 2} b^2+
b(\alpha\partial)^\mu A_\mu
- {1\over 2} M^{\mu\nu}  (V^2)_{\mu\nu}
\ \}
\eea
 For the sake  of notational simplicity, we use the graded commutator  
notation, $[X,Y]^a= f^a_{bc}X^bY^c$, and
$F_{\mu\nu}(A,g)=\partial_\mu A_\nu -\partial_\nu A_\mu+g[A_\mu,A_\nu]$.

 The relation between the renormalization constants $Z$ and the constants $c$
appearing in eq.~(\ref{eq:solution}) is
 \bea Z^{\nu}_{A\mu}  &=& \delta ^\mu_\nu
 (1+{{c_1}\over 2})  
 +c^\nu_{2\mu}
  \CR
Z_c &=& 1+c_3+ {{c_1}\over 2}
\CR
Z_g &=& 1    - {{c_1}\over 2}
\eea
Thus, the effect of renormalisation, constrained by the BRST invariance, 
can be seen as the following redefinitions
of fields and parameters:
 \bea \label{mult}
A_\mu  &\to&          Z^{\nu}_{ A\mu} A_\nu
  \CR
c      &\to&    Z_c  c           \CR
g      &\to& Z_g  g           \CR
K ^\mu  &\to&            Z^{-1 \mu}_{A\nu}  K^\nu \CR
L      &\to&           {1\over {Z^{c   }}} L  \CR
  \bar c &\to&              \bar c    \CR
 b      &\to&              b         \CR
\beta      &\to&                
\beta           \CR
\alpha^{\mu\nu}      &\to&      Z^{-1 \mu}_{A\rho}
 \alpha^{\rho\nu}      \CR
V &\to& V \CR
M &\to &M
\eea 	
Here $A_\mu$ and $K^\mu$ transform contragrediently under renormalization, as
do $c$ and $L$, so that the master equation is invariant
under renormalization in any finite order.

	Equation~(\ref{mult}) shows that the
renormalization is (matricially) multiplicative 
for  the  fields, sources and parameters 
of the theory and that, as compared to the covariant case, the
breaking  of Lorentz invariance by the gauge fixing term
induces a mixing by the renormalization of the
4 components in $A_\mu$ and $K_\mu$. Let us stress that the simplicity of the renormalization of
$\bar c$ $b$, $\beta$ and $\alpha^\mu_\nu$, which generalizes that of
covariant
renormalizable  gauges, is a particularity of linear  gauges for
which one can maintain the $K$ and $\bar c$ dependences through the
combination $K+\alpha\partial \bar c$.  

	These equations indicate the existence of a renormalized BRST symmetry for
the action $\I$, in eq.~(\ref{action}). We will shortly display its
expression. It is however instructive to rederive    the renormalized action 
$\I$, using 
  the  method displayed
in  \cite{Baulieu}, which has the advantage of determining
at the same time 
 the renormalized BRST invariance of the theory.

In this method, one parametrizes
     the renormalized action
$ \Sigma_\rr
$, including  all relevant couterterms, as
  \be\label{Sr}
\Sigma
_\r=S_\r(\Phi,V   )
 +\sum_{                  \Phi}
(
 K_\Phi   ,    s_\r \Phi)  +Ms_\r V
\ee
Recall that $\Phi   $ stand for all fields,  $ A, c, \bar c, b   $.
 One has  assumed 
 that the dependence on the sources   
         $K$'s  of the
BRST transformations is linear, which will be checked by self consistency.
 Then, 
     $s_\r \Phi$ stand for 
field polynomials  in the fields $\Phi$, 
which
 can be 
   expressed as the 
action on
     $\Phi$ 
 of a yet undetermined graded differential operator  $s_\r$.

One can show that   the content of the Ward identities of
the  
BRST symmetry is that (i)  $S_\rr$ is invariant under the action of $s_\r$
and (ii)   
$s_\r$ is a nilpotent operator \cite{Baulieu}: 
\be
s_\r^2=0
\ee
and
\be
s_\r S_\rr=0
\ee

   To compute the possible action of $s_\r \Phi$ , with $s^2_\r\Phi=0$,
one uses
 the results of section~(3).
Up to inessential overall factors, 
 the only freedom    left  in  determining the action of $s_\r$
                  is a matricial redefinition
 of $A_\mu$, that is, $A_\mu \to Z_\mu^\nu A_\nu$, and the rescaling 
$g\to Z_g g$
 so that the requirement
$s^ 2_\r= 0$ implies 
\bea
s_\r V&=&-VV\CR
s_\r c   &=&-{{Z_g g}\over 2}[c,c]-V^\nu_\mu x^\mu\partial_\nu c\CR
 s_\r Z^\nu_\mu A_\nu   &=&
\partial_\mu c+Z_g g  [Z^\nu_\mu A_\nu , c]
-V^\kappa_\lambda x^\lambda \partial_\kappa  Z^\nu_\rho A_\nu
- V_\mu^\rho Z^\nu_\mu A_\nu\CR
\eea
and
\bea
s_\r \bar c   &=&Z_c^{-1}
b\CR
 s_\r b   &=&0.
\eea
This allows one to identify 
$ Z^{\mu}_{A\nu}= Z^\nu_\mu$. Notice also the freedom in rescaling the
field $b$. In the expression of the action, one furthermore sees that a
rescaling of $b$ only amounts to a rescaling of the partition function $Z$,
which is unobservable.

             We also remark parenthetically that one can write
\bea\label{sr}
 s_\r  A_\mu   &=&
 \partial_\mu' c+Z_g g  [  A_\mu , c]
-V^{'\kappa}_\lambda x^{'\lambda} \partial'_\kappa A_\mu
- V_\mu^{'\rho}   A_\mu
\eea
where $\partial'_\mu=Z^{-1\nu}_\mu\partial_\nu$,
$x ^{'\mu}=Z^{ \mu}_\nu x ^\nu$
and $V
^{ ,
\kappa}_\lambda=Z^{ \kappa}_\mu V^\mu_\nu Z^{-1\nu}_\lambda $.
 This shows
another interesting property of the class of 
non-covariant gauges that we have introduced:
 the transformation of the components of $A$ implied by 
breaking of Lorentz invariance, while maintaining
 BRST invariance,
can be absorbed into a
transformations of space time coordinates, $x \to x'$, together 
with redefinitions   of
constant ghost matrix elements  $V\to V'$. (One has             $s_\r
V'=-V'V'$.) 

The non-trivial
part of the cohomology of $s_\r$
with dimension 4 is $|\ \partial _\mu A_\nu^\r -\partial
_\nu A_\mu^\r +Z_g[A_\mu^\r,A_\nu^\r]\ |^2$,
with $A^\r_\mu=Z^\nu_\mu A_\nu$; the rest of $S_\r
$ can only
be $s_\r$-exact terms with dimension~4. By using the anti-ghost
equation of motion as a Ward identity,
which implies that no quartic ghost interactions occur in the action, 
together with the property that the $b$-dependent part of the action does not
need counter-terms, one concludes that $S_\rr$
must be of the form
\bea
S_\rr=\int d^4x \big( \  {1\over 4}
|\ \partial _\mu A_\nu^\r -\partial _\nu A_\mu^\r
+Z_g[A_\mu^\r,A_\nu^\r]\ |^2
+
\CR
+Z_c
 s_\r \{ \bar c   [ \ 
(  \alpha  \partial)^\mu A_\mu^\r +
 {{ \beta\over 2 }} b \ ] \ \} \ \big)
\eea
If we now expand $S_\rr$, using the 
definition  of $s_\r$, and insert this into
eq.~(\ref{Sr}), we exactly recover the formula giving $\I$ in
eq.~(\ref{action}). 

 The renormalized action $\I$, which is
 suitable for the computation of renormalized Green functions,
is thus invariant by construction under the action of the  operation $s_\r$,
which is the renormalized expression of BRST symmetry.

\section{Gauss-BRST Ward Identity}

In the remainder of the article we shall study the Coulomb
gauge limit.  The preceding results hold in particular for interpolating
Coulomb gauges, when the matrix $\alpha$ is diagonal,  and $\alpha_{00}=a$
and
$\alpha_{ii}=1$.  As compared to the case of covariant gauges, there is
just one extra  renormalization constant, with $A_i\to 
Z_{\vec{A}} A_i$ and 
$A_0\to Z_{A_0} A_0$, and $ Z_{\vec{A}} \neq Z_{A_0}$.

	In \cite{Z1} a Gauss-BRST Ward identity was derived in the formal $a = 0$
Coulomb gauge.  This identity is the functional analog of the operator
statement that the BRST symmetry transformation is generated by the
Gauss-BRST charge.  In the present section we shall show that this identity
holds in the Coulomb gauge limit $a \to 0$ from the Landau-Coulomb
interpolating gauge.

	Consider the partition function in the Euclidean theory, with
Coulomb type interpolating gauge, 
$\vec{\nabla} \cdot \vec{A} + a\dot{A}_0 = 0$ (or $= f$).  Having assured
ourselves of Lorentz invariance, we set $V = M = 0$, and the partition
function becomes
\begin{equation}
Z(J, K, L) \equiv \int d\Phi \exp[ - \Sigma + (\Phi, J) ],
\end{equation}
where $\Phi = (A, c, \bar{c}, b)$, and
\be
 \Sigma(\Phi, K, L) 
 \equiv  S_{\rm FP}(\Phi) + (K, sA) + (L, sc). \label{eq:definesigma}
\ee
With $\Sigma = \int d^4x \ \Lambda$, the Lagrangian density reads
\be
\Lambda \equiv (1/4) F_{\mu\nu}^2 
 + (K_\mu + \partial'_\mu \bar{c} ) D(A)_\mu c + L(-g/2) \cdot (c \times c)
-\partial'_\mu b A_\mu + {{\beta \over 2}} b^2 ,
\ee
where
$\partial'_\mu \equiv (\alpha \partial)_\mu 
\equiv (a\partial_0, \vec{\nabla})$.
We are interested in the Coulomb gauge limit $a \rightarrow 0$.  Because
of the gauge arbitrariness of the Coulomb gauge discussed in the
Introduction, this limit may be
$\beta$-dependent, with
$\beta = 0$ for the Landau-Coulomb gauge or $\beta = a$ for
the Feynman-Coulomb gauge. 

	The Lagrangian density is BRST-closed, $s\Lambda = 0$.  This implies the
existence of an identity associated with the corresponding Noether
current, which we now derive. For this purpose we make the infinitesimal
change of variable of integration corresponding to a space-time dependent BRST
transformation
\be
\Phi'_\alpha = \Phi_\alpha + \epsilon(x) s\Phi_\alpha ,
\ee
where $\epsilon(x)$ is space-time dependent, and
$\alpha$ is an index the runs over all components of all integration
variables.  This change of variables leaves the measure $d\Phi$
invariant, and so, because
$\epsilon(x)$ is arbitrary, it yields the identity
\be
0 = \int d\Phi(\partial_\mu j_\mu + sA_\mu J_{A_\mu} + sc J_c + b
J_{\bar{c}})
\exp[ - \Sigma + (\Phi, J) ],
\ee
where $j_\mu$ is the Noether current of the BRST symmetry of $\Lambda$.  If
we integrate this identity over all space-time, the term
$\partial_\mu j_\mu$ is annihilated, and we obtain the Zinn-Justin
equation used previously. Instead we integrate over \mbox{3-space} only, with
spatially periodic boundary conditions, and obtain
\be
\int d^3x (J_{A_\mu} {{\delta Z \over \delta K_\mu}} 
+ J_c {{\delta Z \over \delta L}} 
- J_{\bar{c}} {{\delta Z \over \delta J_b}} )
= \partial_0 \int d\Phi \ Q \  \exp [ - \Sigma + (\Phi, J) ].
\ee

	The conserved BRST charge Q is calculated from
\be
Q = \int d^3x [(sA_\mu) {{\partial \Lambda \over \partial(\partial_0 A_\mu)}}
+ (sc) {{\partial \Lambda \over \partial(\partial_0 c)}}
+ (s\bar{c}) {{\partial \Lambda \over \partial(\partial_0 \bar{c})}} ],
\ee
where the fermionic derivatives are left derivatives, which gives
\be
Q = \int d^3x [ - c D_i F_{0i} - (K_0 + a\partial_0\bar{c})(sc)
+ abD_0 c].
\ee
We wish to express the BRST charge in a way which will provide a
Ward identity satisfied by the quantum effectve action $\Gamma$.  For this
purpose we observe that $Q$ may be written
\be
Q = \int d^3x [ - c {{\delta \Sigma \over \delta A_0}} 
+ K_0 {{\delta \Sigma \over \delta L}} ] + Q_a
\ee
where
\be
Q_a \equiv a \int d^3x \ s(bA_0 - \partial_0\bar{c} c) 
\label{eq:Qsuba}
\ee
is proportional to $a$, and is the integral of a BRST-exact density.

	The quantity ${{\delta \Sigma \over \delta A_0}}$ is the
left-hand side of Gauss's law.  In a canonical formulation, 
it is also the generator of local gauge
transformations, so the first term of $Q$ has the form of the generator of an
infinitesimal gauge transformation with generator $-c(x)$.  For this reason,
the last expression for the BRST charge $Q$ remains correct if coupling to
quarks is included in the Lagrangian density, and also in the phase-space
representation which we shall introduce in the following section.

	From this expression for $Q$ we obtain
\be
\int d^3x (J_{A_\mu} {{\delta Z \over \delta K_\mu}} 
+ J_c {{\delta Z \over \delta L}} 
- J_{\bar{c}} {{\delta Z \over \delta J_b}} )  \CR
= \partial_0 \int d^3x (J_{A_0}{{\delta Z \over \delta J_c}} 
-K_0 {{\delta Z \over \delta L}} ) 
+ Z \langle \dot{Q}_a \rangle,
\ee
The expectation-value $\langle \dot{Q}_a \rangle$ is calculated in the
presence of all sources.  In terms of the generator of connected
correlation functions, $W(J, K, L) = \ln Z(J, K, L)$,
this identity reads
\be
\int d^3x (J_{A_\mu} {{\delta W \over \delta K_\mu}} 
+ J_c {{\delta W \over \delta L}} 
- J_{\bar{c}} {{\delta W \over \delta J_b}} )  \CR
= \partial_0 \int d^3x (J_{A_0}{{\delta W \over \delta J_c}} 
-K_0 {{\delta W \over \delta L}} ) 
+ \langle \dot{Q}_a \rangle.
\ee
We make the Legendre transformation to the quantum effective
action $\Gamma(\Phi, K, L)$, which satisfies
\be
\int d^3x (
{{\delta \Gamma \over \delta A_\mu}} {{\delta \Gamma \over \delta K_\mu}}
+ {{\delta \Gamma \over \delta c}} {{\delta \Gamma \over \delta L}}
+ b {{\delta \Gamma \over \delta \bar{c} }} ) \CR
= \partial_0 \int d^3x (c {{\delta \Gamma \over \delta A_0}}
- K_0 {{\delta \Gamma \over \delta L}} )
- \langle \dot{Q}_a \rangle.
\ee

	Because $Q_a$ is proportional to $a$, one has $Q_a = 0$ in the
formal Coulomb gauge $a = 0$.  However Feynman integrals diverge in the limit
$a \to 0$, so a precise evaluation
is required to determine whether or not
$\langle \dot{Q}_a \rangle$ 
really vanishes in the limit $a \to 0$.  In the
following section we study this limit from the Landau-Coulomb interpolating
gauge,
$\beta = 0$, by means of a phase-space representation.  By power counting
of the $k_0$ integrations, it is found that the correlation functions with
dimensional regularization are finite in the limit $a \to 0$.  It is
found that, although 
$\langle \dot{Q}_a \rangle$ does not in fact vanish linearly with $a$,
nevertheless it does vanish like
\be
\langle \dot{Q}_a \rangle = O(a^{1/2}).
\ee
in the limit $a \to 0$.  (See remark 3 at the end of the following section.)

	We now take the limit $a \to 0$, and set
$\langle \dot{Q}_a \rangle = 0$.  Only first functional derivatives of
$\Gamma$ appear, so the unacceptably singular expression of correlation
functions at coincident points is absent, and this identity imposes a
constraint on the renormalization constants of the elementary fields.
	As before, the Lagrangian multiplier fields $b$ and $\bar{c}$ may be
eliminated by means of their equations of motion, and the Gauss-BRST identity
simplifies to  
\be
\int d^3x 
( {{\delta \tilde{\Gamma} \over \delta A_\mu}} 
{{\delta \tilde{\Gamma} \over \delta K_\mu}} 
+ {{\delta \tilde{\Gamma} \over \delta c}}
{{\delta \tilde{\Gamma} \over \delta L}} )
 = \partial_0 \int d^3x (c {{\delta \tilde{\Gamma} \over \delta A_0}}
- K_0 {{\delta \tilde{\Gamma} \over \delta L}} ).
\ee	

	According to our results on renormalization, the quantum effective action
$\tilde{\Gamma}$ is finite when expressed in terms of renormalized quantities,
\be
\tilde{\Gamma}(X) = \tilde{\Gamma}_r(X_r),
\ee
where $X = (A, c, K, L, g, \Lambda)$ and 
$X_r = (A_r, c_r, K_r, L_r, g_r, \mu)$.  Here $\Lambda$ is the usual
ultraviolet regularization parameter, and $\mu$ is a renormalization mass. 
The renormalization constants satisfy
\be
{Z_K}_\mu^\nu = {Z_A^{-1}}_\mu^\nu \ \ \ \ \ \ Z_L = Z_c^{-1}.
\ee
Moreover, for the Coulomb gauge, by rotational invariance, the
matrix ${Z_A}_\mu^\nu$ is given by 
${Z_A}_\mu^\nu = {\rm diag}(Z_{A_0}, Z_{\vec{A}}, Z_{\vec{A}},
Z_{\vec{A}})$, and 
${Z_K}_\mu^\nu = {\rm diag}(Z_{A_0}^{-1}, Z_{\vec{A}}^{-1}, Z_{\vec{A}}^{-1},
Z_{\vec{A}}^{-1})$.  Consequently the Gauss-BRST identity reads
\be
\int d^3x (
{{\delta \tilde{\Gamma}_r \over \delta A_{r,\mu} }} 
{{\delta \tilde{\Gamma}_r \over \delta K_{r,\mu} }} 
+ {{\delta \tilde{\Gamma}_r \over \delta c_r}}
{{\delta \tilde{\Gamma}_r \over \delta L_r}} )
 = {{Z_c \over Z_{A_0} }}
\partial_0 \int d^3x (c_r {{\delta \tilde{\Gamma}_r \over \delta A_{r,0} }}
- K_{r,0} {{\delta \tilde{\Gamma}_r \over \delta L_r}} ).
\ee	
Since all other quantities in this equation are finite, the ratio 
$Z_c / Z_{A_0}$ must also be finite.  This implies that in the recursive
renormalization procedure described above, the divergent parts of $Z_c$ and
$Z_{A_0}$ are equal in each order $n$.  The iterative renormalization may
be done so the finite parts are also equal in each order, and the equality
\be
Z_{A_0} = Z_c   \label{eq:ZAequalsZC}
\ee
is maintained.

	For this purpose we must show that the renormalized action
$\tilde{\Sigma}_r$ also satisfies the Gauss-BRST identity.  It is instructive
to first verify directly that $\tilde{\Sigma}$ satisfies this identity. 
Indeed by Noether's theorem the variation of $\Sigma$ under the above
space-time dependent BRST transformation is given by
\be
\delta \Sigma = - \int d^4x \epsilon(x) \partial_\mu j_\mu
\ee
where $j_\mu$ is the Noether current.  On the other hand we have
\bea
\delta \Sigma & = &\int d^4x \epsilon(x) s\Phi_i 
{{\delta \Sigma \over \delta \Phi_i }}  \CR
& = & \int d^4x \epsilon(x) 
({{\delta \Sigma \over \delta K_\mu }}{{\delta \Sigma \over \delta A_\mu }}
+ {{\delta \Sigma \over \delta L }}{{\delta \Sigma \over \delta c }}
+ b{{\delta \Sigma \over \delta \bar{c} }} ).
\eea
Since $\epsilon(x)$ is arbitrary, it follows that $\Sigma$ satisfies,
\be
{{\delta \Sigma \over \delta K_\mu }}{{\delta \Sigma \over \delta A_\mu }}
+ {{\delta \Sigma \over \delta L }}{{\delta \Sigma \over \delta c }}
+ b{{\delta \Sigma \over \delta \bar{c} }} 
= - \partial_\mu j_\mu .
\ee
Upon integrating this equation over \mbox{3-space} and using the above
expression for the BRST charge $Q$, we obtain
\be
\int d^3x (
{{\delta \Sigma \over \delta A_\mu}} {{\delta \Sigma \over \delta K_\mu}}
+ {{\delta \Sigma \over \delta c}} {{\delta \Sigma \over \delta L}}
+ b {{\delta \Sigma \over \delta \bar{c} }} ) \CR
= \partial_0 \int d^3x (c {{\delta \Sigma \over \delta A_0}}
- K_0 {{\delta \Sigma \over \delta L}} ) - \dot{Q}_a.
\ee
We now introduce 
$\tilde{\Sigma}(A, c, K, L) = \int d^4x \ \tilde{\Lambda}$, where
\be
\tilde{\Lambda}(A, c, K, L) \equiv (1/4) F_{\mu\nu}^2 
 + K_\mu D(A)_\mu c + L(-g/2) \cdot (c \times c) ,
\ee
so
\be
\Lambda(A, c, \bar{c}, b, K, L) = 
\tilde{\Lambda}(A, c, K + \partial' \bar{c}, L) 
-\partial'_\mu b A_\mu + {{\beta \over 2}} b^2 .
\ee
By the above reasoning we conclude that  $\tilde{\Sigma}$ satisfies the
functional identity
\be
\int d^3x (
{{\delta \tilde{\Sigma} \over \delta A_\mu}}
 {{\delta \tilde{\Sigma} \over \delta K_\mu}}
+ {{\delta \tilde{\Sigma} \over \delta c}} 
{{\delta \tilde{\Sigma} \over \delta L}} )
= \partial_0 \int d^3x (c {{\delta \tilde{\Sigma} \over \delta A_0}}
- K_0 {{\delta \tilde{\Sigma} \over \delta L}} ).
\ee
If one makes the change of variables
\bea
A_\mu & = & Z_{A_\mu} A_{r,\mu}  \CR
K_\mu & = & Z_{A_\mu}^{-1} K_{r,\mu} \CR
c & = & Z_c c_r \CR
L & = & Z_c^{-1} L_r \CR
g & = & Z_g g_r \CR
\tilde{\Sigma}(A, K, c, L, g) & = & \tilde{\Sigma}_r(A_r, K_r, c_r, L_r, g_r),
\eea
with $Z_{A_0} = Z_c$, this identity remains unchanged, so
$\tilde{\Sigma}_r$ satisfies the same functional identity as $\tilde{\Sigma}$.
This is the required condition for recursive renormalization.

	We have taken the limit $a \to 0$ from the Landau-Coulomb
interpolating gauge, 
$\beta = 0$, for which the estimates of the following section hold.  In this
gauge, as noted at the end of sect. (6), the renormalization constant $c_3 =
0$, so
$Z_g Z_c = 1$.  We therefore obtain in the $a \to 0$ limit from the
Landau-Coulomb interpolating gauge
\be
Z_g Z_{A_0} = 1.
\ee
Consequently the field $gA_0$ is invariant under renormalization
\be
gA_0 = g_rA_{r,0},
\ee
as are its correlation functions, including in particular the zero-zero
component of the gluon propagator,
\be
D_{00}(|\vec{x}|, t) = g^2 \langle A_0(|\vec{x}|, t) A_0(0, 0) \rangle.
\ee
This quantity is independent of the cut-off $\Lambda$ and the renormalization
mass $\mu$, and consequently it can depend only on physical masses such as
$\Lambda_{\rm QCD}$.  This holds for the instantaneous part
of $D_{00}(|\vec{x}|, t)$.  However the instantaneous part of 
$D_{00}(|\vec{x}|, t)$ may not be easy to separate uniquely (for example
even in finite orders of perturbation theory), and a more accessible quantity
is
$U(|\vec{x}|) \equiv - \int dt D_{00}(|\vec{x}|, t)$.  It also depends on
physical masses only, as does its fourier transform
$\tilde{U}(|\vec{k}|)$ which is given simply by
$\tilde{U}(|\vec{k}|) = \tilde{D}_{00}(\vec{k}, k_0)|_{k_0 = 0}$.
We write $\tilde{U}(|\vec{k}|) = g_c^2/\vec{k}^2$.  Here 
$g_c = g_c(|\vec{k}|/\Lambda_{QCD})$ is a running coupling constant
defined in the Landau-Coulomb gauge that depends only on $\Lambda_{QCD}$. 
Such a quantity cannot be extracted from the gluon propagator in covariant
gauges.  Indeed to extract it in covariant gauges one must consider the
Wilson loop which involves \mbox{$n$-point} functions of all order~$n$.

\section{Coulomb gauge limit}
\def\a{a}

We now turn
to a more precise analysis of the behaviour of the correlation functions when
the Coulomb-gauge limit $a \to 0$ is taken from the Landau-Coulomb
interpolating gauge, characterized by
$\beta = 0$.  Because the gauge parameter $a$ provides a
rescaling of the time, instantaneous interactions appear as $a$ approaches 0.
        
	Consider the partition function in the Euclidean theory, with
Landau-Coulomb type interpolating gauge, 
$\vec{\nabla} \cdot \vec{A} + a\dot{A}_0 = 0$, 
\begin{eqnarray} 
Z = \int d^4A dc d\bar{c}db \;  
\exp\{\: - \int d^4x \: [ ({1\over 2})( \vec{E}^2 + \vec{B}^2 )
\: + ib (\vec{\nabla} \cdot \vec{A} + a\dot{A}_0) 
\nonumber       \\
+ \, (a \dot{\bar{c}}D_0 c + \vec{\nabla}\bar{c} \cdot \vec{D} c) \:]
\: \},    
\label{eq:Z1}
\end{eqnarray}
where $t = x_0$ represents Euclidean ``time'', 
$E_i \equiv \dot{A}_i - D_iA_0$; $\vec{B} = \vec{B}(\vec{A})$, 
$D_\mu = D_\mu(A)$.  (The i appears in
front of b, because b is here integrated over a real instead of
imaginary contour.)  For
simplicity, we have suppressed all sources, and a summation on color
indices is understood.  

	We use the Gaussian identity
$\exp[(-{1\over 2}) \int d^4x \; \vec{E}^2] 
= \int d^3P \exp[- \int d^4x \; ( i\vec{P} \cdot \vec{E} + ({1\over 2})
\vec{P}^2]$,
to obain the phase-space representation
\begin{equation}
Z = \int d^4A d^3P dc d\bar{c} db \; \exp( - S), 
\end{equation}
where $\vec{A}$ and $\vec{P}$ are canonical variables, and
\begin{eqnarray}
S \equiv \int d^4x \: [ \: i\vec{P} \cdot (\dot{\vec{A}} - \vec{D}A_0 )  
+ ({1\over2})\vec{P}^2 + ({1\over 2})\vec{B}^2 \: 
\nonumber       \\   
+ ib (\vec{\nabla} \cdot \vec{A} + a\dot{A}_0) 
+ \, (a \dot{\bar{c}}D_0 c + \vec{\nabla}\bar{c} \cdot
\vec{D} c) \:] \:.    \label{eq:phaseaction}
\end{eqnarray}
The phase-space action is BRST-invariant, with $\vec{P}$
transforming according to $sP_i^a = f^{abd}P_i^bc^d$.

        We now make a linear change of
field variable in order to diagonalize the gluon propagator, while keeping the
action local.  We pose
\begin{equation}
A_0 = \vec{\nabla}^2 \psi   \label{eq:A0}
\end{equation}
for which $dA_0 = {\rm const} \: d\psi$,
and we shift $\vec{A}$ by
\begin{equation}
\vec{A} = \vec{A}' - a \vec{\nabla} \dot{\psi},  \label{eq:Ai}
\end{equation}
for which $d^3A = d^3A'$.  This simplifies the Lagrange-multiplier
term
\begin{equation}
ib \: (\vec{\nabla} \cdot \vec{A} + a\partial_0 A_0)
= ib \: (\vec{\nabla}\cdot \vec{A}')
\end{equation}
so it imposes the time-independent
constraint $\vec{\nabla} \cdot \vec{A}' = 0$, and we have
\begin{equation}
i\vec{P} \cdot (\dot{\vec{A}} - \vec{D}A_0 ) = 
i\vec{P} \cdot (\dot{\vec{A'}} 
- a\vec{\nabla} \ddot{\psi} - \vec{D}A_0),
\end{equation}
where
$\vec{D} = \vec{D}(\vec{A}) = \vec{D}(\vec{A}' - a\vec{\nabla}\dot{\psi})$. 
We similarly separate $\vec{P}$ into its transverse and longitudinal parts, 
while keeping the action local, by introducing another lagrange multiplier
field by means of the identity,
\begin{eqnarray}
{\rm const} & = & \int d\Omega \; \delta (\vec{\nabla} \cdot \vec{P} 
 + \vec{\nabla}^2 \Omega)    \nonumber \\
& = & \int d\Omega dv \exp[ -i \int d^4x \: v \:
(\vec{\nabla} \cdot \vec{P} + \vec{\nabla}^2 \Omega) ],
\end{eqnarray}
which we insert into the partition function.  We shift $\vec{P}'$ according
to
\begin{equation}
\vec{P} = \vec{P}' - \vec{\nabla} \Omega,   
\end{equation}
under which $d^3P = d^3P'$, so the new Lagrange-multiplier term becomes
\begin{equation}
i \:v \: (\vec{\nabla} \cdot \vec{P} + \vec{\nabla}^2 \Omega)
= i \:v \: (\vec{\nabla} \cdot \vec{P}'),
\end{equation}
and enforces the time-independent constraint 
$\vec{\nabla} \cdot \vec{P}' = 0$.  The field $\Omega$ represents the
color-Coulomb potential. 

	The partition function now reads
\begin{equation}
Z = \int d^3A' d^3P' db dv d\psi d\Omega dc d\bar{c} \; \exp( - S')  
\end{equation}
where 
\begin{eqnarray}
S' \equiv \int d^4x \: [ \: i\vec{P} \cdot (\dot{\vec{A}} - \vec{D}A_0 )
+ ({1\over 2})\vec{P}^2 + ({1\over 2})\vec{B}^2 \:  
+ i v \: \vec{\nabla} \cdot \vec{P}'
 + ib \: \vec{\nabla} \cdot \vec{A}'   \nonumber \\
+ (a \dot{\bar{c}}D_0 c + \vec{\nabla}\bar{c} \cdot \vec{D} c)],
\end{eqnarray}
$\vec{B} = \vec{B}(\vec{A}) = \vec{B}(\vec{A}' - a \vec{\nabla}\dot{\psi})$,
and $\vec{P} = \vec{P}' - \vec{\nabla}\Omega$.  The first term in $S'$ is
given by
\begin{eqnarray}
i\vec{P} \cdot (\dot{\vec{A}} - \vec{D}A_0 )  & = &  
i\vec{P}' \cdot (\dot{\vec{A'}} 
- a\vec{\nabla} \ddot{\psi} - \vec{D}A_0)
  \nonumber \\
& \, & - i\vec{\nabla} \Omega \cdot (\dot{\vec{A'}} 
- a\vec{\nabla} \ddot{\psi} - \vec{D}A_0).
\end{eqnarray} 
To cancel cross terms in $S'$ we shift the Lagrange
multiplier fields,
\begin{eqnarray}
b & = & b' + \dot{\Omega}   \nonumber  \\
v & = & v' - (a \ddot{\psi} + A_0) - i\Omega,
\end{eqnarray}
with $dbdv = db'dv'$, and obtain, after integrating by parts in space and
time and writing $\vec{\nabla}^2 \psi = A_0$,
\begin{eqnarray}
S' = \int d^4x \: [ && i\vec{P}' \cdot
(\dot{\vec{A'}} - g\vec{A} \times A_0) 
+ i( a\dot{\Omega} \dot{A}_0 + \vec{\nabla}\Omega \cdot \vec{D}A_0 ) 
\nonumber \\       \label{eq:defineSprime}
&& + \, ({1\over 2})\vec{P}'^2
+ ({1\over 2})(\vec{\nabla} \Omega)^2
+ ({1\over 2})\vec{B}^2 \:   \nonumber \\
&& + i v' \: \vec{\nabla} \cdot \vec{P}'
 + \: ib' \: \vec{\nabla} \cdot \vec{A}'
+ (a \dot{\bar{c}}D_0 c + \vec{\nabla}\bar{c} \cdot \vec{D} c) \: ].
\end{eqnarray}

	The remainder of this section is an analysis of the action $S'$.  The
Lagrange multiplier fields
$b'$ and $v'$ enforce the time-independent constraints 
$\vec{\nabla} \cdot \vec{A}' = 0$ and $\vec{\nabla} \cdot \vec{P}' = 0$, on
the canonically conjugate variables
$\vec{A}'$ and $\vec{P}'$, and we call these ``the transverse fields''.  The
bose fields
$A_0$ and $\Omega$ form a pair similar to the pair of fermi fields $c$
and $\bar{c}$, and we call this quartet ``the scalar fields''. 

	The corresponding free action
\begin{eqnarray}
S_0 = \int d^4x \: [ && i\vec{P}' \cdot \dot{\vec{A'}} 
+ \, ({1\over 2})\vec{P}'^2
+ ({1\over 2})(\epsilon_{ijk} \nabla_j A'_k)^2  
+ i v' \: \vec{\nabla} \cdot \vec{P}'  
+ \: ib' \: \vec{\nabla} \cdot \vec{A}'  \nonumber \\  
&& + \, i \, ( a\dot{\Omega} \dot{A}_0 
+ \vec{\nabla}\Omega \cdot \vec{\nabla}A_0 ) 
+ ({1\over 2})(\vec{\nabla} \Omega)^2  \nonumber \\
&& + \, (a \dot{\bar{c}}\dot{c} 
+ \vec{\nabla}\bar{c} \cdot \vec{\nabla}c) \:].
\end{eqnarray}
determines the free propagators.  In momentum space the propagators of the
transverse fields are given by
\begin{eqnarray}
D_{A'_iA'_j} & = & (\delta_{ij} - \hat{k}_i \hat{k}_j)
(k_0^2 + \vec{k}^2)^{-1}
\nonumber \\
D_{P'_iP'_j} & = & (\delta_{ij}\vec{k}^2 - k_i k_j)(k_0^2 +
\vec{k}^2)^{-1}
\nonumber \\
D_{P'_iA'_j} & = & ik_0(\delta_{ij} - \hat{k}_i \hat{k}_j)
(k_0^2 + \vec{k}^2)^{-1},
\end{eqnarray}
whereas the propagators of the scalar fields are given by
\begin{eqnarray}
D_{A_0 \Omega} & = & ( ak_0^2 + \vec{k}^2)^{-1} 
\nonumber \\ 
D_{\Omega \Omega} & = & 0 \nonumber \\
D_{A_0A_0} & = & \vec{k}^2 ( ak_0^2 + \vec{k}^2 )^{-2} \nonumber  \\
D_{c \bar{c}} & = & ( ak_0^2 + \vec{k}^2)^{-1} .
\end{eqnarray}
Propagators of the $\psi$ field are
obtained from 
$\psi = (\vec{\nabla}^2)^{-1}A_0$.  The new fields have conveniently
diagonalized the gluon propagator by separating the 3-dimensionally
transverse and scalar parts.  The transverse
propagators have denominators $(k_0^2 + \vec{k}^2)$, whereas the scalar
propagators have denominators $(ak_0^2 + \vec{k}^2)$.  Thus the scalar
fields have a reaction time of order $a^{1/2}$ which is very rapid as 
$a$ aproaches 0.  Consequently it is natural to integrate out, if possible,
the scalar fields and obtain an effective theory for the transverse degrees of
freedom.  

	To study the limit $a \to 0$, we separate the action into 4 terms,
\be
S' = S_r + S_X + S_Y + S_Z.   \label{eq:decomposeS}
\ee
Here $S_r$ is the free action $S_0$ plus all vertices that
are independent of $a$,
\begin{eqnarray}
S_r \equiv \int d^4x \: [ && i\vec{P}' \cdot
(\dot{\vec{A'}} - g\vec{A}' \times A_0)
+ \, ({1\over 2})\vec{P}'^2
+ ({1\over 2})\vec{B}'^2    \nonumber \\
&& + i \, v' \: \vec{\nabla} \cdot \vec{P}'
 + \: ib' \: \vec{\nabla} \cdot \vec{A}' \nonumber \\  
&& + i\,( a\dot{\Omega} \dot{A}_0 + \vec{\nabla}\Omega \cdot \vec{D}'A_0 )
+ ({1\over 2})(\vec{\nabla} \Omega)^2
\nonumber \\
&& + (a \dot{\bar{c}}\dot{c} + \vec{\nabla}\bar{c} \cdot \vec{D}' c) \: ],
\end{eqnarray}
where $\vec{B}' \equiv \vec{B}(\vec{A'})$ and 
$\vec{D}' \equiv \vec{D}(\vec{A'})$.  We shall see that $S_r$ has graphs
that diverge as $a \to 0$, but that they cancel by virtue of an
$r$-invariance.  The term
$S_X$ consists of all vertices with 3 scalar fields and one power of $a$,
\be
S_X = ag \int d^4x 
[ - i \nabla_i \Omega (\nabla_i \dot{\psi} \times A_0)
+ \dot{\bar{c}}(A_0 \times c)  
- \nabla_i \bar{c} (\nabla_i \dot{\psi} \times c) ] .
\ee
The term $S_Y$ consists of a
vertex with one power of $a$,
\be
S_Y = ag i \int d^4x P_i'(\nabla_i \dot{\psi} \times A_0) ] .
\ee
There remains
\be
S_Z = \int d^4x (1/2) [ \vec{B}^2(\vec{A}' - a \vec{\nabla} \dot{\psi})
- \vec{B}^2(\vec{A}') ].
\ee
 	
	We first discuss the theory defined
by $S_r$, temporarily ignoring the vertices $S_X$, $S_Y$ and $S_Z$ that
vanish with $a$.  The action
$S_r$ is at most quadratic in the scalar fields.  Its vertices contain no
powers of
$a$ and no time derivatives, so in momentum space there are no factors of
$k_0$ at the vertices of $S_r$.   Consider a closed loop that consists
entirely of scalar propagators with denominators
$(ak_0^2 + \vec{k}^2)$.  It is controlled by a time scale of order
$a^{1/2}$.  The loop integral on
$k_0$ is effected by the change of variable 
$k_0 = a^{-1/2}k_0'$, which effectively eliminates $a$ from the
denominators, but the volume element of the loop integral changes by
$dk_0 = a^{-1/2}dk_0'$.  We conclude that each closed loop that consists
entirely of scalar propagators and vertices of $S_r$ diverges like
$a^{-1/2}$.  

	Nevertheless the theory defined by $S_r$ is finite as $a \to 0$, as we
now show.  We write
\be
S_r = S_{r,1} + S_{r,2}
\ee
where $S_{r,1}$ consists of all terms that contain only transverse fields
and their Lagrange multipliers,
\begin{eqnarray}
S_{r,1} \equiv \int d^4x \: [ && i\vec{P}' \cdot \dot{\vec{A'}}
+ \, ({1\over 2})\vec{P}'^2
+ ({1\over 2})\vec{B}'^2    \nonumber \\
&& + i \, v' \: \vec{\nabla} \cdot \vec{P}'
 + \: ib' \: \vec{\nabla} \cdot \vec{A}' ].
\end{eqnarray}
It is independent of $a$.  The remainder $S_{r,2}$ also depends on the
scalar fields and on $a$. It is helpful to express $S_{r,2}$ in terms of
the color charge density of the transverse fields
$\rho \equiv gP_i' \times A_i'$, and the Faddeev-Popov
operator 
$M \equiv - a\partial_0^2 - \vec{\nabla} \cdot \vec{D}'$ characteristic of
$S_r$.   We have, in an obvious notation,
\be
S_{r,2} = - i (\rho, A_0) + i(\Omega, M A_0) 
+ (1/2)(\vec{\nabla}\Omega, \vec{\nabla}\Omega)
+ (\bar{c}, M c).
\ee
If one integrates out the ghost
fields $c$ and $\bar{c}$, one obtains the Faddeev-Popov determinant $\det M$.
If one next integrates out $A_0$, one obtains
$\delta(M\Omega - \rho)$, which expresses the form of Gauss's law
appropriate to $S_r$.  Finally the integral on
$d\Omega$ absorbs the Faddeev-Popov determinent
\be
\int d\Omega \det M \ \delta(M\Omega - \rho) 
\exp[-(1/2)(\vec{\nabla}\Omega, \vec{\nabla}\Omega)] 
= {\rm const.} \times \exp( - S_{\rm coul}),
\ee
where
\be
S_{\rm coul} \equiv 
(1/2) (\vec{\nabla}M^{-1}\rho, \vec{\nabla}M^{-1}\rho),
\ee
depends on the transverse fields only.  It represents the
non-local color-Coulomb interaction, regularized however by
the finite value of $a$.  Thus the theory described by the local action 
$S_r = S_{r,1} + S_{r,2}$
that contains the scalar fields is equivalent to the theory with transverse
degrees of freedom only, described by the non-local action 
$S_{r,1} + S_{\rm coul}$.

	Moreover $S_{r,1} + S_{\rm coul}$ at finite $a$ provides a regularized
version of the canonical action,
\be
S_{\rm can} = S_{r,1} + S_{\rm coul}|_{a = 0}.
\ee
The canonical action $S_{\rm can}$ is a function of the canonical
variables which are the transverse fields $\vec{A}'$ and $\vec{P}'$.  It is
obtained by formal canonical quantization in the Coulomb gauge, in which one
solves the constraints to eliminate the so-called unphysical degrees of
freedom.  

	To show that the theory described by the local action $S_r$, or
equivalently by $S_{r,1} + S_{\rm coul}$, is finite in the limit $a \to
0$, we observe that the perturbative expansion of
$S_{\rm coul}$ produces ladder
graphs, in which the instantaneous parts are the horizontal rungs,
corresponding to the instantaneous color-Coulomb interaction.  Since these
ladder graphs do not contain any instantaneous closed loops, they are finite
in the limit $a \to 0$.  To
summarize: in the theory described by $S_r$, each closed loop of bose and
fermi scalars diverges like
$a^{-1/2}$, but they precisely cancel to give a result that is finite as $a
\to 0$.  

	It is helpful to exhibit the cancellation between bosons and fermions in
the theory described by $S_r$ by means of an $r$-symmetry.   We
express the action $S_{r,2}$ in terms of the field 
$\bar{\Omega} \equiv \Omega - M^{-1}\rho$,
\be
S_{r,2} = i(\bar{\Omega}, M A_0) + (\bar{c}, M c) 
+ (1/2)(\vec{\nabla}\bar{\Omega}, \vec{\nabla}\bar{\Omega})
+ (\vec{\nabla}\bar{\Omega}, \vec{\nabla} M^{-1}\rho)
+ S_{\rm coul}.
\ee
Let $r$ be a BRST-type transformation that
acts on the scalar fields according to
\bea
rA_0 & = & c \ \ \ \ \ \ \ \ \ \ \ \  rc = 0  \CR
r\bar{c} & = &  -i \bar{\Omega} \ \ \ \ \ \ r\bar{\Omega} = 0,
\eea
and that annihilates the
transverse fields and their Lagrange multipliers, 
$r\vec{A}' = r\vec{P}' = rb' = rv' = 0$.
It is nil-potent, $r^2 = 0$.  The action $S_{r,2}$ may be written
\be
S_{r,2} = S_{\rm coul} + r \Psi,
\ee
where
\be
\Psi = - \ (\bar{c}, MA_0) 
+ (i/2)(\vec{\nabla}\bar{c}, \vec{\nabla}\Omega')
+ i (\vec{\nabla}\bar{c}, \vec{\nabla}M^{-1}\rho),
\ee
and we have
\be
S_r = S_{r,1} + S_{\rm coul} + r \Psi.
\ee 
The first 2 terms depend on the transverse fields only, and are thus
$r$-invariant, 
$rS_{r,1} = rS_{\rm coul} = 0$.  The last term $r\Psi$, which contains all
the dependence on the scalar fields, is $r$-exact.  We have $rS_r = 0$, and
$r$ is indeed a symmetry of the theory defined by $S_r$.  Now consider the
integral over the scalar fields while the transverse fields and their
Lagrange multipliers are held fixed.  The effective action for the scalar
fields is $r\Psi$, which is
$r$-exact.  A theory whose action is exact under a BRST-type transformation
is called ``topological'', and has the property that the partition function, 
\be
\int dA_0 d\Omega dc d\bar{c} \ \exp(- r\Psi),
\ee
is constant under continuous variations of the external parameters, namely
the transverse fields and the parameter $a$.  We have obtained the previous
result, with the understanding that the cancellation of bose and fermi loops
that diverge in the limit $a \to 0$ is preserved by the $r$-symmetry of
$S_r$.  The
$r$-symmetry which transforms $A_0$ into $c$ explains the equality of
renormalization constants that holds in the limit $a \to 0$,
$Z_{A_0} = Z_c$, which was established in the last section .

	We now come to the remaining vertices of $S'$, namely  $S_X$, $S_Y$ and
$S_Z$.  These vertices formally vanish in the limit $a \to 0$, and they would
not appear in formal canonical quantization in the Coulomb gauge.  However
because, as we have seen, there are closed loops in the expansion of
$S_Z$ that are of order
$a^{-1/2}$ (and that cancel pairwise), we must verify whether insertions into
these loops of the vertices $S_X$, $S_Y$ or $S_Z$ may give a finite result. 
These vertices are not $r$-invariant, so if there are such
contributions there is no reason to expect that they cancel.

	Consider first the vertices of $S_X$ which we call $X$-vertices. 
(Similarly we call $r$-vertices the vertices of $S_r$ etc.)  The
$X$-vertices are linear in $a$.  They also contain one time derivative, so in
momentum space they contain one power of 
$k_0 = a^{-1/2}k_0'$.  Thus overall when an $X$-vertex is inserted into a
closed loop of scalar propagators it gives a contribution of order
$a^{1/2}$.  As we have observed from $dk_0 = a^{-1/2}dk_0'$, the volume
element for a closed loop consisting of scalar propagators is of order
$a^{-1/2}$.  Thus the presence of a single
$X$-vertex in a closed loop of scalar propagators and $r$-vertices would give
a finite limit, except for the fact that such a loop is odd in $k_0'$ at
large 
$k_0'$, and consequently a closed loop of scalar propagators with a single 
$X$-vertex is reduced to order $a^{1/2}$.  By the same reasoning, a closed
loop of scalar propagators that contains two $X$-vertices (and is thus even
in $k_0'$) is also of order $a^{1/2}$.  Thus a single closed loop with
one or two $X$-vertices vanishes like $a^{1/2}$ as $a \to 0$.  However a
closed scalar loop with two $X$-vertices has two external scalar lines,
because each X-vertex is trilinear in the scalar fields.  Consequently such a
loop may be inserted into a closed scalar loop whose remaining vertices are
all
$r$-vertices. [See fig. (1).]  This gives a two-loop graph, with two
$X$-vertices, each of order
$a^{1/2}$, and two closed loops of scalar propagators, each of order
$a^{-1/2}$.  This is finite in the limit $a \to 0$.  (Further insertion of
$X$-vertices gives a vanishing contribution in the limit.)  We conclude that
scalar bose or fermi closed loops do not decouple as 
$a \to 0$, but give a finite
two-loop graph.  This contribution is missing in
formal canonical quantization in the Coulomb gauge.

	The analysis of the vertices of $S_Y$ is
similar.  Each $Y$-vertex contains two scalar fields and one $\vec{P}'$
field. It also contains one power of $a$ and one time derivative, so a
$Y$-vertex is also of order $a^{1/2}$.  Again, insertion of single $Y$-vertex
into a scalar closed loop would be finite except that it is odd in $k_0'$. 
We cannot connect up two
$Y$-vertices by an additional scalar propagator because $Y$-vertices are
bilinear in the scalar fields.  However the $Y$-vertex contains the $\vec{P}'$
field which has the
$P_i'-A_j'$ propagator $P_{ij}(\hat{k})k_0(k_0^2 + \vec{k}^2)^{-1}$ that
contains $k_0$ in the numerator.  (It is the only propagator with $k_0$ in
the numerator.)  Now consider a closed loop that consists of scalar
propagators and one
$\vec{P}'-\vec{A}'$ propagator.  All the vertices are $r$-vertices except
for one  $Y$-vertex at one end of the $P_i'-A_j'$ propagator. [See fig.
(2).]  When the loop momentum
$k_0$ is of order
$a^{-1/2}$, the Y-vertex is of order $a^{1/2}$, the $P_i'-A_j'$
propagator is of order $a^{1/2}$, and the volume element of the loop
integral is of order
$a^{-1/2}$, so overall this closed loop is of order
$a^{1/2}$.  However it has two scalar external lines that emerge from the two
ends of the $\vec{P}'-\vec{A}'$ propagator.  Consequently this closed loop,
which is of order $a^{1/2}$, may be inserted into in a scalar closed loop
consisting of
$S_r$ vertices which is of order $a^{-1/2}$. 
[See fig. (3).]  This again gives a finite two-loop contribution
that is missing in canonical quantization in the formal Coulomb gauge. 
(Further insertions of Y-vertices give a vanishing contribution in the limit.)

	Finally, the vertices of $S_Z$ give vanishing contribution
in the limit $a \to 0$, because when they contain 2 or 3 scalar fields they
also contain 2 or 3 powers of $a$ respectively.

	We summarize the results of this section:  (1) The diagrams for which
the $k_0$ integrations would diverge in the Coulomb-gauge limit, $a \to 0$,
have been  have been shown to cancel at finite $a$.  The
remaining diagrams are finite in this limit by power counting of the $k_0$
integration.   (2) There are two-loop graphs of the scalar particles
$A_0$-$\Omega$ and
$c-\bar{c}$ that are finite in the limit $a \to 0$, and that are missing
from canonical quantization in the formal Coulomb gauge.  It remains a
logical possibility that these graphs are mere gauge artifacts that do not
contribute to a gauge-invariant expectation-value such as a Wilson loop. 
However there is at the moment no argument to show that this is true.
 
{\bf Remarks}

1.  The correlation functions that do not involve the field $P$ are the same
as in the configuration-space representation, so the finiteness of the
unrenormalized correlation functions in the Coulomb-gauge limit of the
Landau-Coulomb interpolating gauge also holds in each order $n$ for the
configuration-space correlation functions.  This implies that the
configuration-space generating functionals
$Z$, $W$, $\Gamma$ and $\tilde{\Gamma}$ are also finite in the limit
$a \to 0$.  Here an ultraviolet dimensional regulator $\epsilon$ is understood
to be in place.  For the diagrams we have examined, the $a$-dependence at
small $a$ is given by $a^{-m/2}$, where $m$ is a non-negative integer.  (The
terms with negative powers of $a^{1/2}$ cancel.)  These powers are
$\epsilon$-independent, and so cause no trouble in the $\epsilon \to 0$ limit
(as would, for example, terms like $a^\epsilon$).  This is because the terms
that diverge with $a$ come from divergences in the one-dimensional $k_0$
integrations and are not affected by dimensional regularization which is a
continuation in the number of spatial dimensions.  Likewise the cancellation
of terms that diverge as $a \to 0$ is assured by $r$-invariance, and is also
dimension-independent.  Moreover
$\tilde{\Gamma}_{\rm div}^n(a, \epsilon)$, eq.~(\ref{eq:solution}), has a
simple pole structure in $\epsilon$.  Consequently the finiteness of 
$\tilde{\Gamma}^n(a,\epsilon) = \tilde{\Gamma}_R^n(a, \epsilon) +
\tilde{\Gamma}_{\rm div}^n(a, \epsilon)$ 
as $a \to 0$ implies that the residue of
$\tilde{\Gamma}_{\rm div}^n(a, \epsilon)$ and
$\tilde{\Gamma}_R^n(a, \epsilon)$ are separately finite as 
$a \to 0$.  Although we have not made an exhaustive
examination of all diagrams, we expect that the remaining diagrams
behave similarly, and thus that the renormalized correlation functions
are finite in the Coulomb-gauge limit of the interpolating Landau-Coulomb
gauge.

2.  We may regard the finite value of the 2-loop scalar graphs that are
missing in the formal Coulomb gauge, $a = 0$, as an anomaly of the
$r$-symmetry; for the action $S'(a)$ is $r$-invariant at $a = 0$,  $rS'(0) =
0$, but not at finite $a$, and the symmetry is not regained
in the limit $a \to 0$.  This comes about because individual graphs diverge
in this limit, and they combine with subgraphs containing $r$-noninvariant
vertices which vanish in the limit, to give a finite result.  However the
divergent graphs result from a part of the action $r\Psi(a)$ that is
$r$-exact at finite $a$, and thus topological.  This assures that the
divergent graphs cancel each other, so that the limit is finite.  It
also preserves the equality, $Z_{A_0} = Z_c$, among the limiting
renormalization constants found in the last section,
eq.~(\ref{eq:ZAequalsZC}), which would hold if the transformation
$rA_0 = c$ were actually a symmetry of the limiting theory.

3.  To establish the Gauss-BRST Ward identity of the last section, there
remains to verify that
$\langle \dot{Q}_a \rangle = 0$ in the limit $a \to 0$, where the expectation
value is calculated in the presence of all sources.  Here
$Q_a$, is the part of the total BRST charge $Q$ defined in
eq.~(\ref{eq:Qsuba}),
\be
Q_a = a \int d^3x [bD_0c - \partial_0bc 
+ \partial_0\bar{c} (-g/2)c \times c].
\ee 
To evaluate $\langle \dot{Q}_a \rangle$, one makes a diagrammatic expansion of
each term by the method of the present section, using 
$b = b' + \dot{\Omega}$.  The only non-zero propagator of the $b'$ field is
the $b'-A'_i$ propagator, 
$k_i(ak_0^2 + \vec{k}^2)^{-1}$.  For example, consider the contribution of
the term 
$\dot{b}^dc^d = \dot{b}'^dc^d + \ddot{\Omega}^dc^d$.  The term
$\ddot{\Omega}^dc^d$ looks dangerous because it contains two powers of $k_0$. 
However the only non-zero propagator of the 
$\Omega$ field is the $\Omega-A_0$ propagator, and the vertex where $A_0$
is absorbed is proportional to $a$.   A typical graph representing the
contribution of $\ddot{\Omega}^dc^d$ to the fourier transform
$\int dt \exp(ip_0t) \langle \dot{Q}_a(t) \rangle$ 
is illustrated in fig 4.  (Note that $Q_a$ is a color scalar, so it must
decay into at least two quanta, namely $c$ and $A_i$ in fig. 4.) This graph
contributes
\bea
ap_0 \ (2\pi)^{-4}\int d^4k   \ k_0^2 (ak_0^2 + \vec{k}^2)^{-1}
 \ a(k_0+q_0) [a(k_0 + q_0)^2 + (\vec{k} + \vec{q})^2]^{-1} \CR
\times  k_i [a(k_0 + p_0)^2 + \vec{k}^2]^{-1}.
\eea
We rescale the variable of integration $dk_0 = a^{-1/2}dk_0'$, and obtain a
contribution of leading order $a^{1/2}$, keeping in mind that terms that
are asymptotically odd in $k_0'$ are suppressed by $a^{1/2}$.  The other terms
are evaluated similarly.  One finds that each term of
$Q_a$ gives a contribution to the expectation-value of order $a^{1/2}$. 
QED.  

4.  We we have seen that a closed loop of unphysical particles, with
propagators $(k_1^2 + k_2^2 + k_3^2 - ak_0^2)^{-1}$, is of order
$a^{-1/2}$ as the Coulomb-gauge limit of the Landau-Coulomb interpolating
gauge is approached.  This behavior came from the rescaling, $k_0 =
a^{-1/2}k_0'$, that makes the loop integral of
order $dk_0 = a^{-1/2}dk_0'$.  However in the
light-front or axial gauge, the unphysical propagator is 
$[a(k_1^2 + k_2^2) + (1+a)(k_3^2 - k_0^2)/2]^{-1}$ or 
$[a(k_1^2 + k_2^2 - k_0^2) + k_3^2 ]^{-1}$, and the required rescaling,
$(k_1, k_2) = a^{-1/2}(k_1', k_2')$ or
$(k_0, k_1, k_2) = a^{-1/2}(k_0', k_1', k_2')$, gives an uphysical closed loop
integral of order $a^{-1}$ or $a^{-3/2}$.  Thus the light-cone and axial gauge
limits appear to be more singular than the Coulomb-gauge limit, and additional
cancellations would be required to give finite correlation functions.

\section{Conclusion}

	We briefly review our results.  We have addressed the problem of the
existence of ``physical gauges'', by the device of interpolating gauges which
interpolate linearly between a covariant gauge, such as
the Feynman or Landau gauge and a physical gauge such as the Coulomb or
light-cone gauge.  For example, the interpolating Landau-Coulomb
interpolating gauge is defined by the gauge condition 
$a\partial_0A_0 + \grad \cdot \vec{A} = 0$, which gives the Landau gauge for
$a = 1$, and the Coulomb gauge is achieved in the singular limit $a \to 0$. 
More generally an interpolating gauge is defined by the condition 
$\alpha^{\mu\nu}\partial_\nu A_\mu = 0$ (or $= f$), where $\alpha$ is a
non-singular numerical matrix, and a ``physical'' gauge is a limiting case
in which $\alpha$ becomes singular. 

	In general the interpolating gauge
breaks Lorentz invariance as well as local gauge invariance.  Nevertheless we
are able to establish the existence of the perturbative expansion and
perturbative renormalizability of the interpolating gauges in full generality,
by extending the BRST method to include the Lorentz group in addition to the
usual local gauge group.  This extension is necessary to control the form of
divergences, for example to show that the divergent coefficients of the
term $c_E \vec{E}^2 + c_B \vec{B}^2$ are equal, $c_E = c_B$.  The enumeration
of the possible divergence terms that are BRST-invariant is not substantially
more difficult than for Lorentz-covariant gauges.  Moreover the matrix
$\alpha$ is a gauge-parameter in the sense that the expectation values of
physical observables are independent of $\alpha$, as long as $\alpha$ is
non-singular.  Thus the interpolating gauges are strictly equivalent to the
covariant gauges.  

	However the singular limit to a physical gauge is quite subtle.  It is
analyzed in the present article for the Coulomb gauge limit, $a \to 0$,
from the Landau-Coulomb interpolating gauge.  There are closed bose and
fermi-ghost loops that become instantaneous in the limit $a \to 0$ and that
individually diverge like $a^{-1/2}$.  We we use a phase space
representation and a linear shift of field variables to exhibit the
cancellation of loops that diverge like $a^{-1/2}$, and to show by power
counting of the $k_0$ integrals that this limit gives finite correlation
functions. 

	An important aspect of this limit is that there are also closed bose and
fermi one-loop graphs that are not present in the formal Coulomb gauge ($a
= 0$).  Although they vanish like $a^{1/2}$, they cannot be neglected because,
when these one-loop graphs are inserted into the above-mentioned closed loops
that diverge like
$a^{-1/2}$, they give a finite contribution.  Consequently the closed bose and
fermi-ghost loops do not decouple in the Landau-Coulomb gauge limit, but give
a finite two-loop contribution.  
 	
	One logical possibility is that these two-loop ghost contributions are
merely a gauge artifact that do not actually contribute to expectation-values
of gauge-invariant objects such as Wilson loops.  However there is at present
no argument in hand to show this.  If these two-loop bose and
fermi-ghost graphs do contribute to physical expectation values, then the
traditional picture of the Coulomb gauge would have to
be revised.  The state space would not be simply describable in terms
of transverse gluons.  In
the latter case the Coulomb gauge is not more unitary than other gauges, in
the sense that it cannot be simply described in terms of the classical
dynamical variables that remain after the constraints are solved.  Indeed we
are unable to provide a set of Feynman rules to be used in the Coulomb
gauge at $a = 0$, although we have shown that both the unrenormalized and
renormalized correlation functions are finite in the limit $a \to 0$ of
the Landau-Coulomb interpolating gauge.

	Nevertheless there is a reward to be gained by taking this limit.  For we
have shown that the Gauss-BRST Ward identity holds in the Coulomb gauge limit
of the Landau-Coulomb interpolating gauge.  This identity is the functional
analog of the operator statement that the BRST symmetry transformation is
generated by the Gauss-BRST charge.  Among other things, it
implies that $gA_0$ is invariant under renormalization,
$gA_0 = g^{(r)} A_0^{(r)}$.  This means that all correlation functions of
$gA_0$ are renormalization-group invariants, including in particular the
time-time component of the gluon propagator 
$g^2D_{00}$.  It depends only on physical masses such as $\Lambda_{QCD}$,
but is independent of the cut-off or the renormalizaion mass.  Thus the
Coulomb-gauge limit of the Landau-Coulomb gauge provides direct access to
renormalization-group invariant quantities, whereas no component of the gluon
propagator has this property in covariant gauges.  Indeed in covariant gauges
one must go to the Wilson loop, which involves gluon correlation functions of
all orders, to obtain a renormalization-group invariant quantity.  For this
reason the Coulomb gauge may prove advantageous for non-perturbative
formulations.  In particular, the instantaneous part of 
$g^2D_{00}$ may be a confining color-Coulomb potential that may serve as
an order parameter for confinement of color \cite{Z1}.

Acknowledgements: We wish to thank the Newton Institute  for its
hospitality, where part of this work was done.  One of us (LB) wishes to thank
SISSA.  One of us (DZ) recalls with pleasure stimulating conversations with
Massimo Porrati and Martin Schaden.

\section{Figure Captions}

\begin{enumerate}
\item Diagram with 2 scalar loops and 2 $X$-vertices, and any number of
$r$-vertices.
\item  One-loop diagram with a single $P'-A'$ propagator and an $r$-vertex.
\item  Insertion of graph of fig. 2 into a closed scalar loop of
$r$-vertices.
\item  A typical graph contributing to 
$\langle \dot{Q}_a \rangle$.
\end{enumerate}

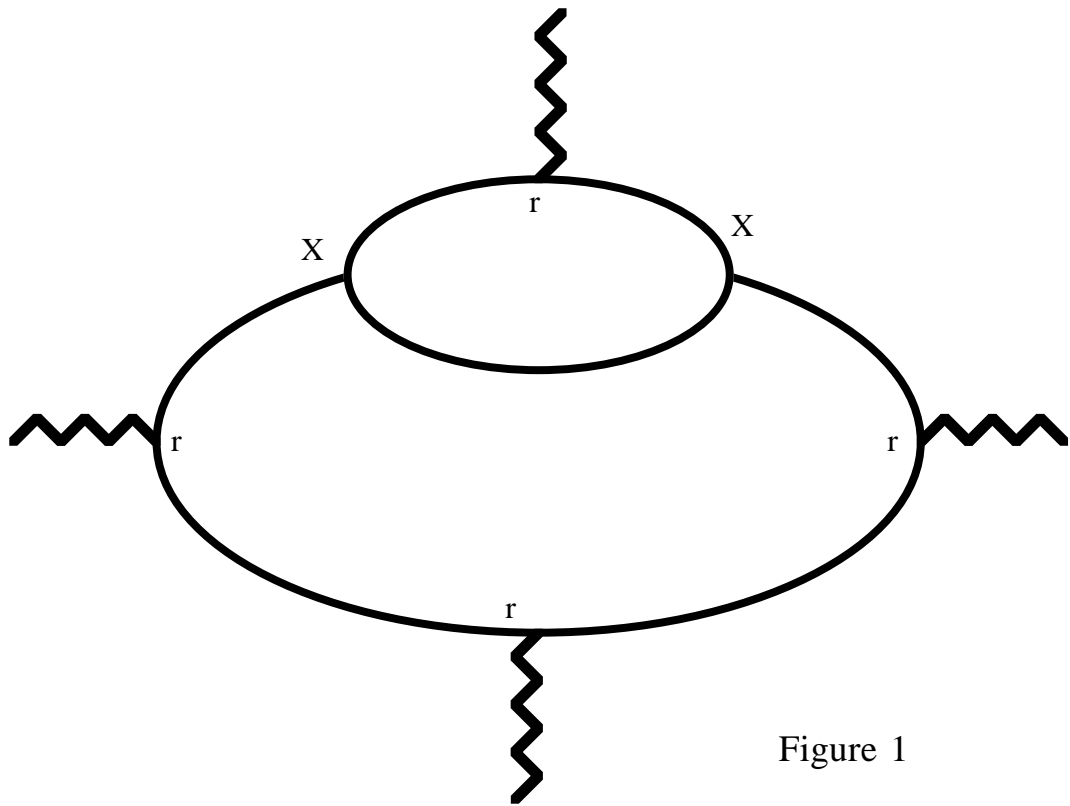
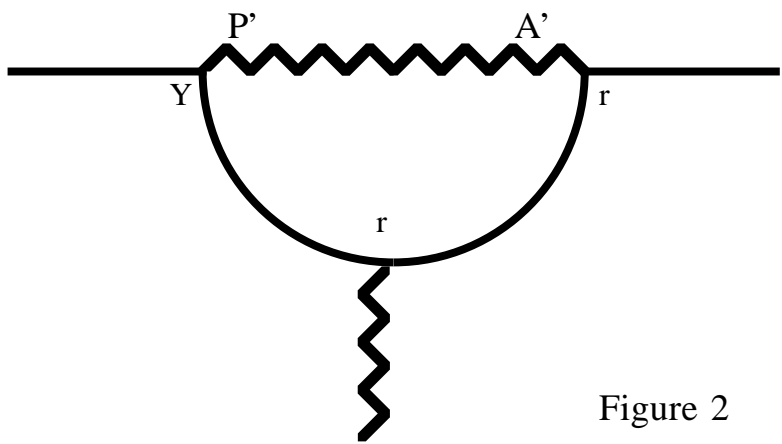
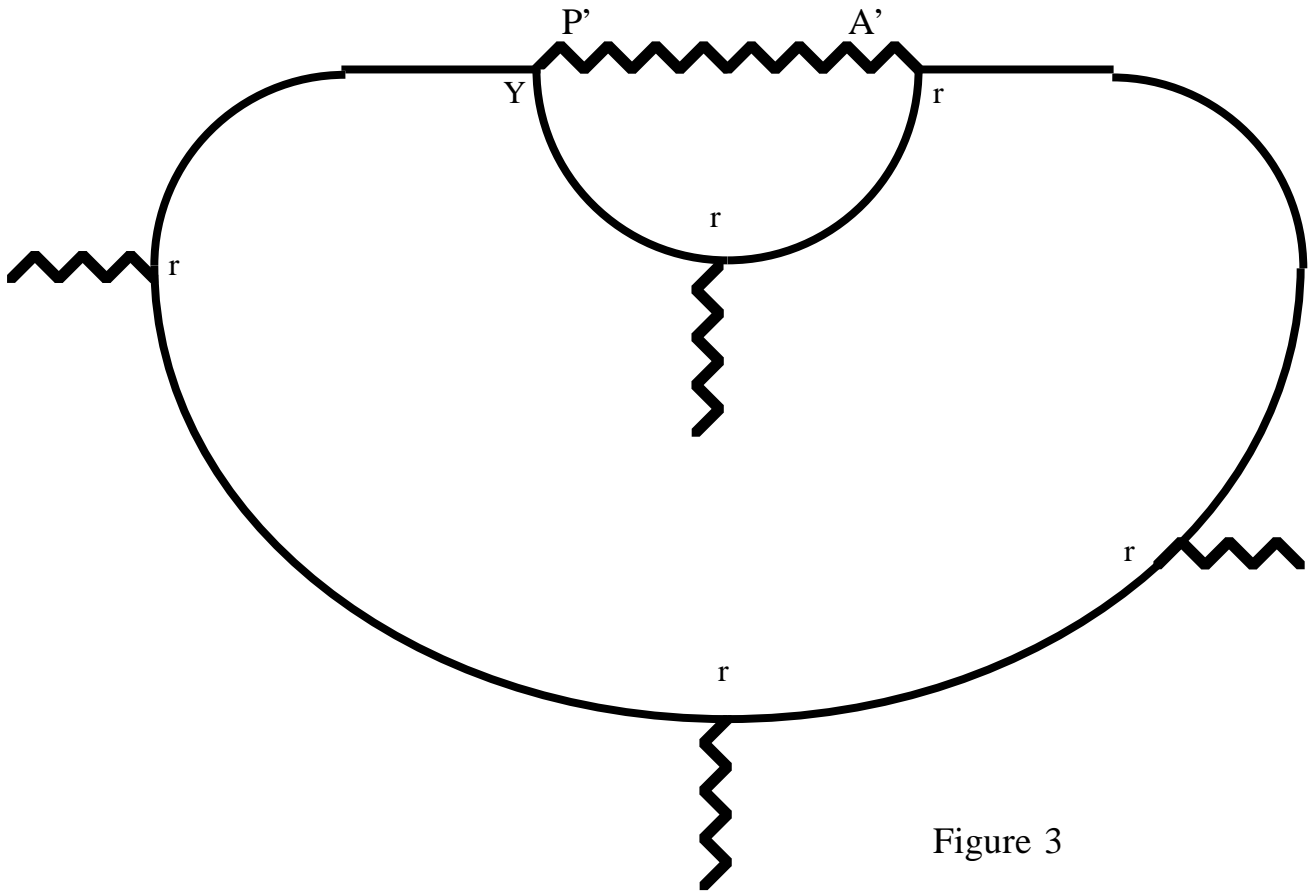
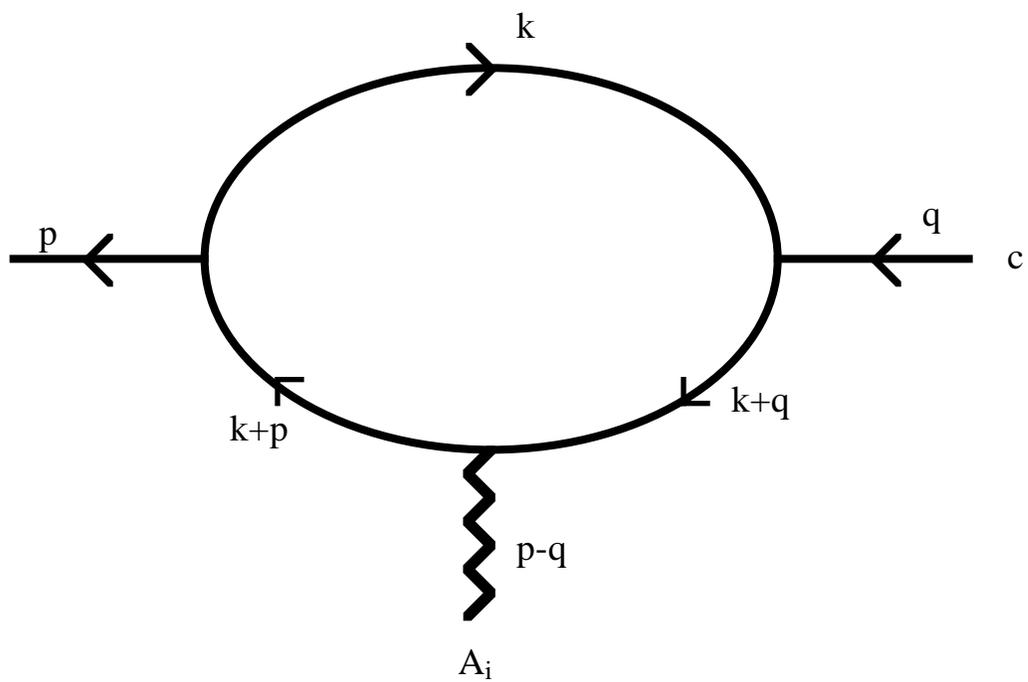
\end{document}